\documentstyle[12pt,twocolumn,floats,aps,epsf]{revtex}
\begin{document}
\onecolumn
\draft
\tighten

\setlength{\textwidth}{16.0cm}
\newcommand{\beqn}{\begin{eqnarray}}
\newcommand{\eeqn}{\end{eqnarray}}
\newcommand{\lwf}[1]{\mbox{\large $#1$}}

\title{Halo Excitation of $^6$He in Inelastic and
Charge-Exchange Reactions}
\author{
\bf Russian-Nordic-British Theory (RNBT) collaboration \\ }
\author{
S.N.Ershov\thanks{Permanent address: JINR, Dubna, Russia},
T.Rogde, \\ }
\address{
{\small SENTEF, Department of Physics,
University of Bergen, Norway} \\ }
\author{
B.V.Danilin\thanks{Permanent address: RRC The Kurchatov Institute,
 Moscow, Russia}, J.S.Vaagen, \\ }
\address{
{\small NORDITA, Copenhagen, Denmark and SENTEF, Department of Physics,
University of Bergen, Norway} \\ }
\author{
I.J.Thompson \\ }
\address{
{\small  Department of Physics, University of Surrey, Guildford, UK} }
\author{
 F.A.Gareev \\ }
\address{
{\small JINR, Dubna, Russia}\\ }

\maketitle
\widetext

\begin{abstract}
Four-body distorted wave theory appropriate for  nucleon-nucleus reactions
leading to 3-body continuum excitations of two-neutron Borromean halo nuclei is 
developed. The peculiarities of the halo  bound state and 3-body continuum 
are fully taken into account by using the method of hyperspherical harmonics.
The procedure is applied for $A=6$ test-bench nuclei; thus we report
detailed studies of inclusive cross sections for
inelastic $^6$He(p,p$'$)$^6$He$^*$ and charge-exchange $^6$Li(n,p)$^6$He$^*$ 
reactions at nucleon energy 50 MeV. The theoretical low-energy
spectra exhibit two resonance-like structures. The first (narrow) 
is the excitation of the well-known $2^+$ three-body resonance. 
The second (broad) bump
is a composition of overlapping soft modes of multipolarities 
$1^-, 2^+, 1^+, 0^+$ whose relative weights depend on transferred
momentum and reaction type. Inelastic scattering
is the most selective tool for studying the soft dipole excitation 
mode.
\end{abstract}


\vspace{1cm}
\centerline{Submitted to Phys. Rev. C on May 1, 1997}
\centerline{Preprints: NORDITA 97/18N; Surrey CNP-97/6; }
\centerline{LANL preprint nucl-th/9705002 }

\onecolumn
\section{Introduction}

Recent success in developing experimental methods for dripline nuclei, that in
particular allow exploration of halo phenomena in light nuclei, has put
on the agenda a need for appropriate theoretical methods which take
into account the peculiarities of weakly bound and spatially extended systems.
For Borromean two-neutron halo nuclei ($^6$He, $^{11}$Li, etc.) an understanding
of the essential halo  structure has been obtained in the framework of 3-body 
models \cite{prep93}.
Reactions involving these nuclei present however, at least a 4-body problem.
The direct solution of 4-body systems is extremely difficult, and
 approximate methods are required.  For high
energy elastic scattering and  relativistic fragmentation of Borromean halo
nuclei, a 4-body Glauber method has been developed \cite {Gl4Suz,Gl4IJT}.
For Coulomb breakup or electromagnetic dissociation (EMD) the
first order (Alder-Winther) perturbation theory or an equivalent semiclassical
treatment \cite {beba88} has been used, but with exact 3-body continuum wave functions 
\cite{boris93,ferr93,li11c}.
Also for the (anti)neutrino induced reactions on $^6$Li populating
the $^6$He and $^6$Be
3-body continua, have proper final state wave functions recently been 
used \cite{shul93}.

The most reliable information on  properties of halo nuclei, especially
for the low-lying part of excitation spectra, is experimentally obtainable by 
intermediate energy  elastic and inelastic scattering and 
charge-exchange reactions. The distorted wave theory is the most common way to
analyse such processes \cite{Satch}, but for halo systems 
their spatial granularity as well as peculiarities of their quantum 
structure have to be taken into account. The 3-body interaction
dynamics defines the low-lying part of excitation spectra, in particular the
soft modes of Borromean systems, and has to be treated properly. Until now,
only the hyperspherical harmonics (HH) method \cite{dancon} is able to
provide a formulation of the scattering theory to the 3-body Borromean 
continuum. 
The Faddeev equations technique \cite{gloc} has been developed to investigate 
breakup of 3-nucleon systems, but has hitherto not been 
applied  to investigate the continuum in Borromean nuclei. The coordinate 
complex rotation method \cite {csoto,Ikeda} gives Gamow and
not scattering states, and is mostly suitable for searching for
resonance positions or poles in the complex energy plane.

 In previous studies \cite{prep93,PhR}, the HH
method gave a very successful and comprehensive
description of  data on weak and electromagnetic characteristics of the $^6$He
and $^6$Li systems, and of the absolute values of differential cross sections
of (p,n), (p,p$'$) and (n,p) reactions to the bound states of $A = 6$
nuclei. These nuclei still represent the best testbench for quantitative 
calculations
for Borromean halo nuclei. In the present work we develop distorted wave
theory for inelastic and charge-exchange reactions leading to the 3-body
continuum. For the continuum excitations of $^6$He
we perform a detailed analysis of inclusive excitation and differential
cross-sections for beam energies in the range of the GANIL facility, where
such experiments are in progress. Investigations of continuum spectra of $^6$He
are also subject of future experiments at Kurchatov Institute
(Moscow), NSCL (Michigan), RIKEN (Tokyo), JINR (Dubna) and GSI (Darmstadt).

\section{Short Preamble}

The known spectrum of $^6$He contained until  quite recently only the $0^+$
bound state, the well known $2^+$ (1.8 MeV) 3-body resonance and then a desert 
in the 3-body $\alpha$+n+n continuum up to the $^3$H + $^3$H
threshold at about 13 MeV \cite{aj1}. With radioactive nuclear beam techniques 
and dynamic approaches to 3-body continuum theory \cite{dancon} new 
possibilities are opened, and we may now ask 
 to what extent our knowledge of $^6$He is complete, and what specific
influence the halo has on the continuum structure. The so-called soft dipole
mode suggested in \cite{hans88,suz91} was the first example of this quest.
That it is not a simple binary core-point dineutron resonance in
$^{11}$Li  seems now
widely accepted, nor is it in  $^6$He according to our recent calculations
\cite{he6ex}. Is it rather a genuine 3-body resonance
or just a dynamically induced very large dipole moment, or the consequence of
two-body final state interactions?
Soft modes of other multipolarity   have also been theoretically
 suggested \cite{fayans}. Their presence  needs clarification, both theoretical
and experimental.

The most natural way to observe soft modes in exotic nuclei is
by inelastic excitation of  radioactive beams. In the $^6$He case,
however, only results of
fragmentation experiments without reconstruction of inclusive spectra have
been published \cite{tanihata,koba92,balamuth}. Other ways  include
transfer reactions (like $^7$Li(n,d)$^6$He at $E_n$ = 56.3 MeV \cite{Bradynd})
or charge-exchange reactions of (n,p) type on $^6$Li. At $E_n$ = 60 MeV the 
$^6$Li(n,p)$^6$He reactions have been measured, but with poor statistics and
limited angles \cite{Bradynp}. In heavy-ion charge-exchange reactions
$^6$Li($^7$Li,$^7$Be)$^6$He a broad bump at excitation energy $\sim$ 6 MeV was
observed \cite{Sakuta,Janec}, but different assignments were made about
its multipolarity.

In our recent work \cite{he6ex,he6cont} we have used  several methods to
investigate   the internal structure of the 3-body continuum, 
as well as the transition properties for accessible nuclear reactions 
in terms of nuclear and
electromagnetic response functions. {\it Our methods have the advantage
that, even off a resonance, the continuum
structure can still be investigated while taking into account all final
state interactions.}

We have predicted \cite{he6ex} surprising richness of the
$^6$He continuum structure, and applied  elaborate methods to explore the
nature of different modes of excitation of the Borromean halo continuum
\cite{he6cont}.
In 3-body dynamics we have found two kinds of phenomena. First we
have the true 3-body resonances with characteristic 3-body phase shifts
\cite{dancon}
crossing  $\pi /2$, and with resonance behaviour in all partial waves. 
Secondly, we find  structures (exhibiting fast growth of 3-body
phase shifts up to $\pi/2$ in some partial waves) which are often caused by  
resonant and/or virtual states in binary subsystems.  
We call these structures 3-body virtual excitations.
The analysis of the 
lowest partial components (those having the physically most transparent 
meaning) enables us in the HH method to obtain comprehensive insights into 
both the 3-body effects, and into the influence of resonances in binary 
subsystems on 3-body amplitudes.

In \cite{he6ex} we predicted and in  \cite{he6cont} explored in detail, 
in addition to the well-known $2^+$ resonance, (i)
a second $2^+$ and a $1^+$ resonance in the $^6$He
continuum which both qualify as 3-body 
resonances, (ii) a soft dipole mode which does not but is a 3-body virtual
excitation, and (iii) unnatural parity modes.  Because of the halo 
structure of the ground state, there are peaks in the isoscalar responses of 
the soft monopole mode and soft dipole mode, even though there 
are no resonances in the low-energy continuum region.  
A higher-energy ``breathing mode'' appears as well, in the 
monopole continuum.

Summarising the extended analysis  of \cite{he6cont},
we show in Table~\ref{table3} the positions and widths of possible
resonances obtained by different methods. 
\begin{table}[hbt]
\begin{tabular}{|c|cc|cc|cc|cc|}
  &  HH &  & CS1\cite{csoto}& &
    CS2\cite{Ikeda} & & Exp.\cite{aj1}& \\
\hline
 $J^{\pi}$&  $E$ & $\Gamma$ &$E$& $\Gamma$ &$E$& $\Gamma$ &$E$& $\Gamma$   \\
\hline
$0^+_1$  & 0 & & 0 & & 0 &  & 0&   \\
$2^+_1 $ &  1.72 & 0.04 & 1.71 & 0.06 & 1.77 & 0.26 & 1.8 & 0.113  \\
$2^+_2 $ &  4.0 & 1.2 & - & - & 3.5 & 4.7 & - & -  \\
$1^- $ &  not & found & not & found  & not & found & - & -  \\
$1^+ $ &  4.4 & 1.8 & - & - & 4.0 & 6.4 & - & -  \\
$0^+_2 $ &  6.0 & 6.0 & - & - & 5.0 & 9.4 & - & -  \\
\end{tabular}
\caption{Comparison of resonance positions and widths of $^6$He.
Results from the present Hyperspherical Harmonics Method (HH) and the
Complex Scaling Method (CS) \protect\cite{csoto,Ikeda} are shown, together with
experimental data. Resonance positions are given relative to the theoretical 
ground state.} 
\label{table3}
\end{table}
All of them give about the same
positions, but different widths which should be testable experimentally.

The way in which these structures could be
experimentally observed depends on the reaction considered, as will now be
demonstrated via nucleon charge-exchange and inelastic scattering to the 
$^6$He continuum.
A complicating feature is the overlapping of resonances and soft modes in 
the region of excitation energies 3--6 MeV, thus only a more detailed analysis
of excitation functions and angular distributions may possibly distinguish 
those states. This will be a central issue of this paper, where we will
show that even inclusive cross sections will be informative.


\section{Model description}
\label{model}

In this section details are given on the physical ingredients of the model
we have developed for calculating inelastic and charge-exchange reactions to
low-energy continuum states in $^6$He. Structure and reaction scenarios
are often intertwined in a very complicated way. In some situations, such as the
very dilute matter of halo nuclei, the reaction dynamics becomes simpler, and
an approximate scheme using distorted waves  is  reasonable.
In the DW framework, as follows from formulae below, the reaction amplitude
has three ingredients:

  A. The structural information contained in the
 transition densities which describe the response of the nuclear system to an
 external field,
 
  B. The effective interactions between projectile and target
  nucleons, and 
  
  C. The distorted waves describing the relative motion of
  projectile (ejectile) and target (residual) nucleus.

\subsection  {Nuclear structure }

   For description of the nuclear structure  we have
used the 3-body $\alpha+N+N$ model. In this model, the total wave function
is represented as a product of wave functions describing the internal structure
of the $\alpha$-core and the relative motion of three interacting constituents 
(see appendix~A). The method of hyperspherical harmonics (HH) (see
Refs.~\cite{prep93,dancon,danden}) was used to solve the  Schr\"{o}dinger 
3-body equation, for both bound and continuum states.  A modified SBB
Gaussian type $\alpha$$N$ interaction \cite{sack54} with purely repulsive
$s$-wave component (Pauli core) \cite{prep93} and the ``realistic'' 
GPT $N$$N$ interaction \cite{GPT} were used.

We are now going to apply this model to continuum low-energy excitations above 
the (3-body) breakup
threshold. The main model assumption about the factorization of the wave
function into two parts suggests that low-lying nuclear transitions of interest 
are connected with
excitations of the two valence particles in the halo outside the
$\alpha$-core. This assumption is physically reasonable for the low energy
spectrum since $\alpha$-core excitations must involve a significant
energy transfer due to the particularly stable structure of the 
$\alpha$-particle.

The $^6$He continuum reveals a variety of structures:
the  $1^+$ spin-flip resonance has an almost pure shell-model structure 
($p_{3/2} p_{1/2}$); 
the $2^+_1$ and $2^+_2$ resonances are on the other hand of strongly mixed 
nature; there are 3-body virtual excitation such as the soft dipole 
and monopole modes.

For qualitative insight in the resonance structure, 
tables \protect\ref{he6-0+}-\ref{he6-2+} give partial wave function norms 
in the interior region
$\rho_0 < 15$ fm for all resonances and the $1^-$ peak both in $LS$
and (Jacobi) $jj$ coupling scheme. These norms reflect the ``eigen'' resonance
properties of any few-body system and measure the continuum strength 
accumulated in the strong interaction and centrifugal barrier regions.   
 A hyperradial resonance wave function in the interior
($\rho < \rho_0$) can be represented in a factorised form:
 \beqn \label{intnorm1}
 \chi (\rho; E) \sim A(E) \cdot
 \Psi^R (\rho) = { c \over {E - (E_{0}-i \Gamma_0/2)}} \cdot  
 \Psi^R (\rho)
\eeqn
where $\Psi^R (\rho)$ has structure similar to that of a bound state.
 
 This general energy dependence is revealed by the reaction cross-section, but 
 for wide resonances it is strongly influenced by the reaction mechanism.


\begin{table}[ht]
\begin{tabular}{|ccccc|c|c|c|c|c|}
 \multicolumn{5}{|c|}{ }
                            &$0^+_1$&  $0^+_2$  &        &$0^+_1$& $ 0^+_2$  \\
$K$ &$L$ &$S$ &$l_x$ &$l_y$ &  g.s. & resonance & $(jj)$ &  g.s. & resonance \\
 \hline
  0 & 0  & 0  &   0  &   0  &  4.   & 15. & $p_{3/2}$ $p_{3/2}$ & 86. & 17. \\
  2 & 0  & 0  &   0  &   0  & 78.   & 30. & $p_{1/2}$ $p_{1/2}$ & 5.  & 68. \\
  2 & 1  & 1  &   1  &   1  & 15.   & 51. & $s_{1/2}$ $s_{1/2}$ & 7.  &  3. \\
\end{tabular}
\caption{Weights of the main components of interior parts of $0^{+}$ wave 
functions of $^6$He in $LS$ and $jj$ representation (ground state and
$ 0^+_2$  resonance at 5 MeV above the breakup threshold).}
\label{he6-0+}
\end{table}


\begin{table}[ht]
\begin{tabular}{|cccc|c|c||c|c|c|}
\multicolumn{4}{|c|}{ }& $ 2^+_1$ & $ 2^+_2$ &Config.& $ 2^+_1$ & $ 2^+_2$ \\
$L$ &$S$ &$l_x$ &$l_y$ &resonance &resonance &$(jj)$ &resonance &resonance \\
 \hline
 1  &  1 &   1  &   1  &    32.   &    58.   &$p_{3/2}$ $p_{3/2}$&33.&45.  \\
 2  &  0 &   0  &   2  &    45.   &    30.   &$p_{1/2}$ $p_{3/2}$&32.&32.5 \\
 2  &  0 &   2  &   0  &    22.   &    11.   &$s_{1/2}$ $d_{5/2}$&21.&13.  \\
    &    &      &      &          &          &$s_{1/2}$ $d_{3/2}$&14.& 8.5 \\
\end{tabular}
\caption{Weights of the main components of the interior $2^+$ resonance
state wave functions of $^6$He in $LS$ and $jj$ representations at 0.8 and 
3.0 MeV above the breakup threshold.
}
\label{he6-2+}
\end{table}

The HH method is particularly suited for Borromean systems due to their simple
asymptotic behavior. The physical characteristics of bound and low continuum
states are concentrated in only a few wave function
components corresponding to the lowest angular momenta and energy configurations
 of the 3-body system. This, combined with a convergence behaviour for 
ground state and resonances which is very much the same, preserves their 
relative  position and enables us to
avoid time consuming calculations. Thus we only take into account the
hyperharmonics with hypermoments $K \le 6$ that correspond to excitation 
energy up to
about 10 MeV $(\kappa \rho \sim K)$. 
Only the specific 3-body virtual nature of the soft dipole mode demands an
substantially larger series
of hyperharmonics to achieve convergence. So we shall use the main components
keeping in mind that the dipole mode will be somewhat shifted to lower
energy within the peak width.

\subsection  {Effective interactions between projectile and target
  nucleons}

The effective NN interaction $V_{pt}$ between projectile and target
nucleons is a key point in the microscopic approaches to the description of
 one-step reactions. It differs from a free interaction
since one of the nucleons is embedded in the nuclear medium, its motion being
restricted by Pauli blocking and interactions with the nuclear environment.
Usually these modifications are expressed by means of density dependence of the
effective interactions, and a lot of work has gone into calculating 
effective interactions starting from the free one. As a rule, the calculations
are based on nuclear matter with applications made to finite nuclei via local
density approximations. These procedures include some uncertainties, which are
especially troublesome when we deal with the lightest nuclei. Some physical
situations exist, however, where the interaction dynamics  simplifies,
and  simpler approaches can be used. At intermediate
energies the impulse approximation has proven to be a very successful.
In this situation
the nucleon-nucleon collision energy is sufficiently large compared to
binding energies, and the modification of the free interaction is not very
significant. Using as effective interaction the free nucleon-nucleon
$t$-matrix, that takes into account an infinite number of rescatterings between
two nucleons interacting via a free NN potential, we obtain a complex,
energy dependent interaction with parameters that can be extracted from
analysis of experimental data on free NN scattering.

 Another simplified
situation occurs in interactions with halo particles, because the halo
particles have  small
binding energy and large probability to be outside the core of strongly
bound nucleons. In the course of interaction
with halo nucleons, small momentum and energy
transfers are not blocked as is the case for interaction with
core nucleons. As a result, the interaction with a halo nucleon is  very
similar to the interaction between two free nucleons, and we can use the
free $t$-matrix interactions, in close analogy with
the impulse approximation at intermediate energies. This approach is in the
spirit of our model for nuclear structure, when the ``active'' part of the
nuclear wave function is defined by the motion of halo particles. In concrete
calculations of inelastic and charge-exchange, we use the $t$-matrix
parametrization of Love $\&$ Franey \cite{LoFr} with  central, tensor and
spin-orbit components. The contribution of an exchange knock-out
amplitude is taken into account in the pseudopotential approximation
\cite{petr}.

To be consistent with the 3-body $\alpha+N+N$ model of nuclear structure,
 we have also to take into
account the $\alpha$N-interaction between the projectile nucleon and $\alpha$-core.
For charge-exchange reactions at low excitation energy,
only the halo nucleons are the active particles in our model. Charge-exchange
with core nucleons must destroy the $\alpha$-core and involve a large
excitation energy, and was therefore neglected in our calculations.
In inelastic scattering the $\alpha$N-interaction can give a contribution also
at low excitation energy. If the $\alpha$-core is taken to have an infinite mass this
contribution will be exactly zero, due to the orthogonality between initial
bound and final continuum states of the 3-body system. In real situations the
center of mass of the total nucleus is somewhat shifted from the 
$\alpha$-particle center of
mass, and we have a finite contribution from this interaction. We
 expect that the role of direct $\alpha$N-interaction  increases
with increasing  transferred momentum. At this stage of our investigations,
we neglect these contributions. Therefore, our calculations
of inelastic scattering should be reasonable only for moderate transferred
momentum. Physically this corresponds to the situation where the $\alpha$-particle
is a spectator, and experimentally it would be realized if only events with
$\alpha$-particles in forward direction were detected. It is necessary
to underline that when we take into account the interaction with halo nucleons,
the recoil effects are treated in an exact way because the wave functions which
we use  are defined in terms of the translational invariant Jacobi coordinates.

\subsection  {Distorted waves}

 For calculations of distorted waves we need to know the optical potentials
describing the nucleon elastic scattering.  For calculations of distorted waves we used a phenomenological
optical potential \cite{phenopt} describing proton elastic scattering from
$^6$Li at energy $E_p$ = 49.5 MeV. The same optical potentials were used in
both incident and exit channels.


\section{Reaction formalism}
\medskip
\par
\begin{figure}[hbt]
\vspace*{1cm}
\centering\leavevmode
 \epsfxsize 12 cm \epsfbox{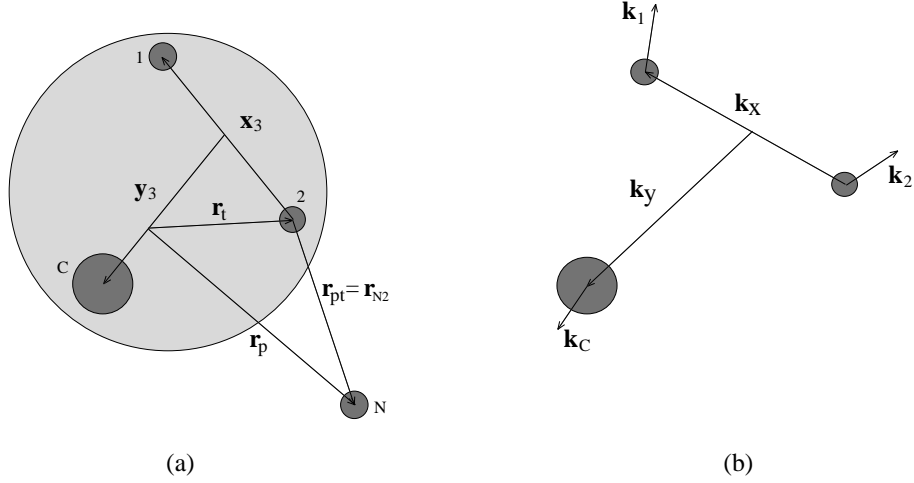}
\vspace*{1cm}
\caption{Spatial coordinates (a) in nucleon-nucleus scattering, and 
particle momenta (b) in the two-neutron halo system.}
\label{fig1}
\end{figure}

Among the  quasielastic reactions, the nucleon-nucleus inelastic
scattering and charge-exchange reactions are the  simplest and best 
understood. Experimental possibilities are now available for applying
these  reactions (in inverse kinematics on nucleon targets) to investigation
 of the structure of exotic halo nuclei.
 The cross section of quasielastic reactions,
$$ N + A \to N'+ C + n_1+n_2  $$ between a nucleon and a two-nucleon halo nucleus
(core $C$, with $A=C + n_1+n_2$ in the g.s.) exciting the latter to the continuum, can be written in the form

\begin{equation}
\sigma =  {(2\pi)^4 \over \hbar v_i} \sum \int d\bbox{k}_N' d\bbox{k}_1
d\bbox{k}_2 d\bbox{k}_{C} \delta(E_f - E_i) \delta(\bbox{P}_f-\bbox{P}_i)
{\mid {\lwf{T}}_{fi} \mid}^2
\end{equation}

\noindent
where $E_i = \varepsilon_N + \varepsilon_A$,
$E_f = \varepsilon_N'+\varepsilon_1+\varepsilon_2+\varepsilon_{C}+Q$,
$\bbox{P}_i = \bbox{k}_N+\bbox{k}_A$,
$\bbox{P}_f = \bbox{k}_N'+\bbox{k}_1+\bbox{k}_2+\bbox{k}_{C}$ are the total
energies and momenta of all particles before and after collisions. In these
expressions Q is the binding energy of the target nucleus in the case of inelastic
scattering, while it is  the difference of binding energies of parent and daughter
nuclei for a charge-exchange reaction. The relative incident velocity is
${\displaystyle v_i = {\hbar k_i \over \mu_i} }$, and 
${\displaystyle \mu_i = {m_N M_A \over m_N+M_A} }$ 
is the reduced mass of the particles before collision. We will work in the center 
of mass (CM) coordinate frame
($\bbox{P}_i$ = 0, $\bbox{k}_N$ = -$\bbox{k}_A$ = $\bbox{k}_i$ ), and use Jacobi
coordinates for particles both in initial and final systems. The coordinates
used are defined on Fig.~\ref{fig1} and are given by

\begin{eqnarray}
\bbox{k}_x &=& \mu_x \Bigl( {\bbox{k}_1 \over m_1}-{\bbox{k}_2 \over m_2} \Bigr),
\ \mu_x = {m_1 m_2 \over m_1+m_2} ,\nonumber \\
\bbox{k}_y &=& \mu_y \Bigl({\bbox{k}_{C} \over m_{C}} -
{\bbox{k}_1 + \bbox{k}_2 \over m_1 + m_2} \Bigr)
,
\ \mu_y = {(m_1+m_2) m_{C} \over m_1+m_2+m_{C} } , \label{eq:defk}\\
\bbox{K} &=& \bbox{k}_1 + \bbox{k}_2 + \bbox{k}_{C} = - \bbox{k}_N'
 = - \bbox{k}_f. \nonumber
\end{eqnarray}

\noindent

In the CM frame $E_f$ = $\varepsilon_f + E_{\kappa} + Q$, where 
$\varepsilon_f$ = ${\displaystyle {\hbar^2 k_f^2 \over 2\mu_f} }$, and 
$\mu_f$ = ${\displaystyle {m_N' M_A \over m_N' + M_A}}$ 
is the reduced mass for the exit channel, while $E_{\kappa}$ = 
$\varepsilon_x+\varepsilon_y$ = 
${\displaystyle {\hbar^2 k_x^2 \over 2\mu_x} + {\hbar^2 k_y^2 \over 2\mu_y}}$ 
is the excitation energy measured from the breakup threshold. Taking into 
account conservation of energy and momenta, the exclusive cross section (when
energies and momenta of all particles are observed) is an average over
initial and a sum over final spin orientation, and can be written as

\begin{equation}
{d^5 \sigma \over d\Omega_x d\Omega_y d\Omega_f d\varepsilon_y dE_{\kappa} } =
(2\pi)^4 {\mu_i \mu_f \over \hbar^4} {k_f \over k_i}
2({\mu_x \mu_y\over \hbar^4})^{3 \over 2}
\sqrt{\varepsilon_y (E_{\kappa}-\varepsilon_y)} {1 \over 2(2J_i+1)}
\sum \mid \lwf{T}_{fi} \mid^2 .\label{eq:exccrs}
\end{equation}

\noindent
The factor $\sqrt{\varepsilon_y (E_{\kappa}-\varepsilon_y)}$ describes the
distribution of energy between different modes of particle motion, and
reflects the phase-space accessible for breakup. The matrix element
$T_{fi}$ includes all the interaction dynamics, and is given in the
Distorted Wave (DW) framework by

\begin{equation}
\lwf{T}_{fi} = <\lwf{\chi}_{f,M_b}^{(-)}(\bbox{k}_f),\lwf{\Psi}^{(-)}_{m_1, m_2, m_C}
 (\bbox{k}_x, \bbox{k}_y) \mid \sum_{t} V_{pt} \mid
\lwf{\Psi}_{J_i M_i},\lwf{\chi}_{i,M_a}^{(+)}(\bbox{k}_i)>, \label{tmatr}
\end{equation}

\noindent
where $\lwf{\chi}_{i,f}^{(\pm)}$ are distorted waves describing relative motion
of colliding nuclei, $\lwf{\Psi}_{J_i M_i}$ and
$\lwf{\Psi}^{(-)}_{m_1, m_2, m_C} (\bbox{k}_x, \bbox{k}_y)$
are the initial bound and final continuum nuclear states, respectively.
Spin projections $M_a$, $M_i$, $M_b$, $m_1$, $m_2$ and $m_C$, together
with relative momenta $\bbox{k}_{i,f}$ and $\bbox{k}_{x,y}$ characterize the
asymptotic state of all particles taking part in the reaction. Continuum wave
functions are a matrix in spin space, and contain the probability for
spin-flip in the course of scattering. Since we are
interested in studying the $^6$He nucleus where the core is an 
$\alpha$-particle with  spin zero, for simplification of notation we omit 
everywhere below mentioning of the core spin projection $m_C$.

In eq. (\ref{tmatr}) $V_{pt}$ is a local, effective nucleon-nucleon 
interaction between projectile and target  nucleons $p$ and $t$, expressed in terms 
of central, spin-orbit and tensor components,

\begin{equation}
V(\bbox{r}_{pt} , \bbox{p}_{pt} ) = \sum_{T} { \left\{ \sum_{S}
t_{ST}^{C}(r_{pt})\sigma_{p}^{S}\cdot \sigma_{t}^{S} +
t_{LS}^{T}(r_{pt})\bbox{L}\cdot\bbox{S} +
t_{T}^{T}(r_{pt})S_{pt}(\bbox{\hat{r}}_{pt}) 
\right\} } \tau_{p}^{T}\cdot\tau_{t}^{T}
\end{equation}

\noindent
where  $\bbox{r}_{pt} = \bbox{r}_{p}-\bbox{r}_{t}$ and 
$\bbox{p}_{pt} = {1\over2}(\bbox{p}_{p}-\bbox{p}_{t})$ are relative distance and
momentum (wave number) between two nucleons, 
$\bbox{S} = {1\over2}(\bbox{\sigma}_{p}+\bbox{\sigma}_{t})$ and
$\bbox{L} = \bbox{r}_{pt}\times\bbox{p}_{pt}$ are operators of total spin and 
orbital angular momentum of the two nucleons, 
$\bbox{p}_{i} = -i \bbox{\nabla_{i}}$,  
$\sigma_{i}^{S} = 1$, $\bbox{\sigma}_i$ for $S$=0, 1 respectively.
$S_{pt}(\bbox{\hat{r}}_{pt})$ is the tensor operator 

\begin{eqnarray}
S_{pt}(\bbox{\hat{r}}_{pt}) &=& { 3(\bbox{\sigma}_{p}\cdot\bbox{r}_{pt})
(\bbox{\sigma}_{t}\cdot\bbox{r}_{pt}) \over r_{pt}^{2} } -
(\bbox{\sigma}_{p}\cdot\bbox{\sigma}_{t}).
\end{eqnarray}

\noindent
Using the method of hyperspherical harmonics, the nuclear wave function above 
the breakup threshold can be written as follows (for details see Appendix~A):

\begin{eqnarray}
\lwf{\Psi}_{m_1, m_2}^{(+)} &=& \sum_{\gamma, J_f, M_f, M_{L_f}}
<s_1 m_1\ s_2 m_2 \mid S_f M_{S_f}> <L_f M_{L_f} S_f M_{S_f} \mid J_f M_f> \label{eq:3scatt} \\
&\times& {\cal Y}^{l_{x}l_y \ \mbox{\large $*$}}_{K_f L_f M_{L_f}}(\Omega_{5}^{\kappa})\ 
\lwf{\Psi}_{\gamma, J_f, M_f} (\bbox{x}, \bbox{y}, \kappa)  
\nonumber
\end{eqnarray}

\noindent
where $\gamma$ is an abbreviation for a set of quantum numbers
$\gamma = \{K_f, L_f, S_f, l_x, l_y \}$, which characterizes 
the relative motion of the three particles flying apart. The continuum wave 
function
$\lwf{\Psi}_{\gamma, J_f, M_f} (\bbox{x}, \bbox{y}, \kappa)$ 
depends on the quantum numbers $\gamma$, Jacobian space coordinates 
$(\bbox{x},\bbox{y})$, 
nuclear excitation energy $E_{\kappa}$ (expressed by the hypermomentum $\kappa$), and 
the total angular momentum $J_f$ and its projections $M_f$,

\begin{equation}
\lwf{\Psi}_{\gamma, J_f, M_f} (\bbox{x}, \bbox{y}, \kappa) = 
{\displaystyle {1 \over (\kappa\rho)^{5/2}}\ \sum_{\gamma', M'_L}
\lwf{\chi}_{K'l'_xl'_y,K_fl_xl_y}^{L'S',L_fS_f}(\kappa\rho)\
\Upsilon^{l'_{x}l'_y}_{J_fK'L'S'M_f}(\Omega_{5})}
\end{equation}

\noindent
The transition amplitude $T_{fi}$ can be further decomposed according 
to eq.~(\ref{eq:3scatt}),

\begin{equation}
\lwf{T}_{fi} =\!  \sum_{\gamma, M_{L_f}, M_{S_f} }\! 
<s_1 m_1 s_2 m_2 \mid S_f M_{S_f} >\
<L_f M_{L_f} S_f M_{S_f} \mid J_f M_f >\
{\cal Y}^{l_{x}l_y \ \mbox{\large $*$}}_{K_f L_f M_{L_f}}(\Omega_{5}^{\kappa})
\ \lwf{T}_{\gamma} 
(\bbox{k}_x, \bbox{k}_y, \kappa ) \label{Tfi3}
\end{equation}         

\noindent
where $T_{\gamma} (\bbox{k}_x, \bbox{k}_y, \kappa )$ formally now has  the same 
structure as any two-body amplitude for excitation of a nuclear state with total
momentum $J_f, M_f$, excitation energy $E_{\kappa}$, and a fixed state of 
relative motion of breakup fragments defined by the quantum numbers $\gamma$:

\begin{eqnarray}
\lwf{T}_{\gamma} (\bbox{k}_x, \bbox{k}_y, \kappa ) &=&          
 <\lwf{\chi}_{f,M_b}^{(-)}(\bbox{k}_f),\lwf{\Psi}_{\gamma, J_f, M_f} (\kappa)
 \mid  \sum_{t} V_{pt} \mid \lwf{\Psi}_{J_i M_i},
 \lwf{\chi}_{i,M_a}^{(+)}(\bbox{k}_i)> \nonumber \\
&=& \sum_{j m}\ <J_i M_i\ j m \mid J_f M_f >\ \lwf{T}_{\gamma, j}^{M_a, M_b, m}
(\bbox{k}_x, \bbox{k}_y, \kappa ) \label{Tfi2}
\end{eqnarray}

\noindent
To calculate the reaction amplitude we use the partial wave decomposition
for the distorted waves $\lwf{\chi}^{(+)}_{i , M_{a}} (\bbox{k}_{i})$,
describing the relative motion of the projectile nucleon and the center 
of mass of the target nucleus:

\begin{eqnarray}
 \lwf{\chi}^{(+)}_{i , M_a} (\bbox{k}_i) &=& \sum_{M_a' } 
 \lwf{\chi}^{(+)}_{i, M_a' M_a} (\bbox{k}_{i}, \bbox{r}_{p}) \vert 
 S_a M_a' \rangle \nonumber \\
 &=& {4\pi \over k_{i} r_{p} }
\sum_{l_a j_a}\ <l_a m_{l_a} S_a M_a \vert j_a m_a>\  i^{l_a}\
Y_{l_a m_{l_a}}^{\mbox{\large $*$}} (\bbox{\hat{k}}_i)\ \lwf{\chi}_{l_a j_a}(k_i,r_p)\
\vert j_a m_a \rangle 
\end{eqnarray}

\noindent
where $\vert j_a m_a\rangle = 
 \sum\ <l_a m_{l_a}' S_a M'_a \vert j_a m_a>\ Y_{l_a m_{l_a}'}
 (\bbox{\hat{r}}_p)\ \vert S_a M'_a \rangle $, and 
$\vert S_a M'_a \rangle $ is the projectile spin function. 
Nuclear formfactors can be defined as follows

\begin{eqnarray}
\lefteqn{ 
< j_b m_b, J_f M_f \mid \sum_{t} V_{pt} \mid J_i M_i, j_a m_a >\ = \sum_{lsj}
<J_i M_i jm \mid J_f M_f>  }\\           
& & \times \imath^{-l}\ { (-1)^{l+s-j+m} \over \hat{s} }\
<j_b m_b \mid [Y_l (\bbox{\hat{r}}_p) \otimes \sigma^s_p ]_{jm} \mid j_a m_a > 
F_{j_b j_a}^{lsj} (\kappa, r_p, {\partial \over \partial r_p } ) 
\label{NuclForm} 
\end{eqnarray}

\noindent
where $\hat{s} = \sqrt{2 s + 1}$.  In the case of nuclear excitations of normal
and unnatural parity the explicit formulas for radial formfactor 
$ F_{j_b j_a}^{lsj}$ are given in \cite{Obsor} and Appendix~B.
Taking into account  these definitions the reaction amplitude 
$T_{\gamma, j}^{M_a, M_b, m} (\bbox{k}_x, \bbox{k}_y, \kappa )$ can be 
written in usual form

\begin{eqnarray}
\lefteqn{\lwf{T}_{\gamma, j}^{M_a, M_b, m} (\bbox{k}_x, \bbox{k}_y, \kappa )  
 =  {(4\pi)^2 \over k_i k_f}
\sum_{l_a j_a l_b j_b l s} i^{l_a-l_b-l}\ Y_{l_a m_{l_a}}^{\mbox{\large $*$}} 
(\bbox{\hat{k}}_i)\ Y_{l_b m_{l_b}} (\bbox{\hat{k}}_f)\  
\lwf{I}_{l_a j_a, l_b j_b}^{lsj}\   
{\hat{j} \hat{j}_b \sqrt{2} \hat{l}_b \hat{l} \over \sqrt{4\pi}} }
\nonumber \\
&\times & <l_a m_{l_a} S_a M_a \vert j_a m_a>\ 
<l_b m_{l_b} S_b M_b \vert j_b m_b>\ <j_b m_b jm \vert j_a m_a>\ 
<l_b 0 l 0 \vert l_a 0>\ 
{ \left\{ \begin{array}{ccc}
l_b & S_b & j_b \\
l   &  s  & j     \\
l_a & S_a & j_a
\end{array} \right\} }
\end{eqnarray}

\noindent
where the radial integrals $I_{l_a j_a, l_b j_b}^{lsj}$ are defined as
follows 

\begin{equation}
\lwf{I}_{l_a j_a, l_b j_b}^{lsj} = \int_{0}^{\infty} dr_p\ r_p^2\
{1\over r_p}\ \lwf{\chi}_{l_b j_b} (k_f,r_p)\
F^{lsj}_{j_a j_b} (\kappa, r_p,{\partial \over \partial r_p} )\
{1\over r_p}\ \lwf{\chi}_{l_a j_a} (k_i, r_p)
\end{equation}

It is useful to compare the expression (\ref{Tfi3}) for the breakup amplitude
$T_{fi}$ with the amplitude for a usual two-body reaction which has the same 
structure as amplitude (\ref{Tfi2}). In the 3-body case, the amplitude
$T_{fi}$ has additional degrees of freedom
which are manifested as dependence on angles $\Omega_5^{\kappa} = \{ \alpha,
\bbox{\hat{k}}_x, \bbox{\hat{k}}_y \}$, where 
$\sin^2 \alpha = {\displaystyle {\varepsilon_x \over E_{\kappa}} }$.
In contrast to a traditional two-body approach, the exclusive cross
section (proportional to $\mid T_{fi} \mid^2$) contains an incoherent sum
over total spin $S_f$ but a coherent sum over total transferred $j$ and
final $J_f$. Consequently, we expect that the
exclusive cross section will be especially sensitive to the correlations in
the nuclear structure. 

The different exclusive cross
sections and correlation distributions will be considered elsewhere, here
we restrict ourselves to inclusive cross sections. To calculate the 
double-differential inclusive cross sections when experiments
measure the energy and angle for one particle, we must integrate
the fivefold exclusive cross section (\ref{eq:exccrs}) over the unobserved
coordinates of breakup particles (angles $\bbox{\hat{x}} = \Omega_x$, 
$\bbox{\hat{y}} = \Omega_y$), and over
various distributions of relative energy $\varepsilon_y$ between fragments:

\begin{equation}
{d^2 \sigma \over d\Omega_f dE_{\kappa} } =
\int_{0}^{E_{\kappa}} d\varepsilon_y \int d\Omega_x d\Omega_y
{d^5 \sigma \over d\Omega_x d\Omega_y d\Omega_f d\varepsilon_y dE_{\kappa} }
\end{equation}

\noindent
Using the following orthogonality properties of the hyperspherical harmonics,

\begin{eqnarray}
\int_{0}^{E_{\kappa}} d\varepsilon_y  \sqrt{\varepsilon_y 
(E_{\kappa}-\varepsilon_y)}
\int d\Omega_x d\Omega_y {\cal Y}_{K' L' M_L'}^{l_x' l_y' \mbox{\large $*$}}
(\Omega_5^{\kappa})
 {\cal Y}_{K L M_L}^{l_x l_y}(\Omega_5^{\kappa}) &  \nonumber \\
 = \ 2 E_{\kappa}^2 \delta_{K' K}
 \delta_{L' L} \delta_{M_L' M_L} \delta_{l_x' l_x}\delta_{l_y' l_y}, &
\end{eqnarray}

\noindent
the inclusive cross section now becomes an incoherent sum over total 
transferred $j$ and final $J_f$ angular momenta, and different 
$\gamma$-components of the final target state are excited independently 
of each other. Thus

\begin{eqnarray}
{d^2 \sigma \over d\Omega_f dE_{\kappa} } &=&
(2\pi)^4 {\mu_i \mu_f \over \hbar^4} {k_f \over k_i}
\sum_{j, J_f, \gamma} {(2J_f+1) \over (2J_i+1)(2j+1)} \nonumber \\
& \times& {1 \over 2}
\sum_{m m_a m_b} \mid \lwf{T}_{\gamma,j}^{m_a m_b m}(\bbox{k}_f, \bbox{k}_i, \kappa)
\mid^2 4 E_{\kappa}^2 ({\mu_x \mu_y \over \hbar^4})^{{3 \over 2}}. 
\label{eq:inccrs}
\end{eqnarray}

\noindent
In this expression the factor $E_{\kappa}^2$, which originates from the 
3-body phase volume, guarantees the correct cross section
behavior at the breakup threshold. From (\ref{eq:inccrs}) it also follows
that, due to the averaging procedure, we lose information about correlations
in relative motion of the breakup particles (which were defined by
${\cal Y}_{K L M_L}^{l_x l_y}(\Omega_5^{\kappa})$ hyperharmonics);
remnants of the complex dynamics that governs the particles motion
are kept only in different shapes and strengths with which various
components of final states are distributed over excitation energies.
One may hope that in the differential inclusive cross sections,
due to specifics of reaction mechanisms, we can under certain conditions
enhance the excitations of some of the components and thus still
obtain valuable information about structures of halo nuclei.

The inclusive excitation cross section
 can be obtained by integrating over all ejectile angles $\Omega_f$:

\begin{equation}
{d\sigma \over dE_{\kappa} } = \int d\Omega_f
{d^2 \sigma \over d\Omega_f dE_{\kappa} }. \label{eq:inclusive}
\end{equation}

\noindent
This cross section describes the distribution of total strength of different
excitation modes over energy spectra in quasielastic reactions.

\section{Results}

With the model described above, we have calculated the excitation
(eq. \ref{eq:inclusive}) and double-differential (eq. \ref{eq:inccrs}) 
inclusive
spectra in the CM system for the charge-exchange reaction $^6$Li(n,p)$^6$He and the
inelastic scattering $^6$He(p,p$'$)$^6$He at $E_N$ = 50 MeV, with
excitation of different $J_f^{\pi_f} = 0^{\pm},1^{\pm},2^{\pm},3^{\pm}$
low-energy states of $^6$He. The corresponding cross sections are shown in
Figs.~\ref{fig3}-\ref{fig6}. In the figures $E^*$ is the nuclear 
excitation energy measured from $^6$He ground state. For inelastic
scattering the initial target state of $^6$He has $J_i$ = 0 and the
total  $j$ transferred has a unique value and coincides
with $J_f$ of the final state. For charge-exchange on $^6$Li
the situation is  more complex: since $J_i$ = 1 for the $^6$Li ground state
it is possible to excite final states of $^6$He with definite
$J_f$ by  different $j$ transfers. All values of $j$
allowed by angular momentum conservation ($\bbox{J}_i-\bbox{J}_f = \bbox{j}$) were
taken into account in our calculations. It follows from equation
(\ref{eq:inccrs}) that every $j$ gives an independent contribution to the
inclusive cross sections.

Our main goal is to demonstrate, that even in the simplest inclusive 
experiments it is still possible to extract information about
structures in the continuum by detailed examination of both excitation and
differential cross-sections.

\subsection  {Two test cases for the model}

Two cases were used to check the model and consistency of
our reaction continuum calculations. 
The sharp $2^+_1 $ resonance at 1.8 MeV was used in the first.
This resonance resembles a usual bound state and can be described 
with good accuracy by calculating it with a boundary condition under 
the barrier corresponding to a discrete state. We use this to calculate 
the differential cross section ${\displaystyle {d\sigma \over d\Omega} }$. 
Next we calculate the double-differential 
${\displaystyle {d^2\sigma \over d\Omega dE^*}}$ 
cross sections for the $2^+_1$ resonance at different $E^*$ and after that
integrate over $E^*$ across the resonance. In fact, calculating a resonance 
width and cross section at peak position, energy integration
has been done analytically since the resonance has the Breit-Wigner form 
(we checked it). In both calculations we got the same results for 
${\displaystyle {d\sigma \over d\Omega} }$.       

\begin{figure}[hbt]
\vspace*{-3cm}
\centering\leavevmode
 \epsfxsize 17 cm \epsfbox{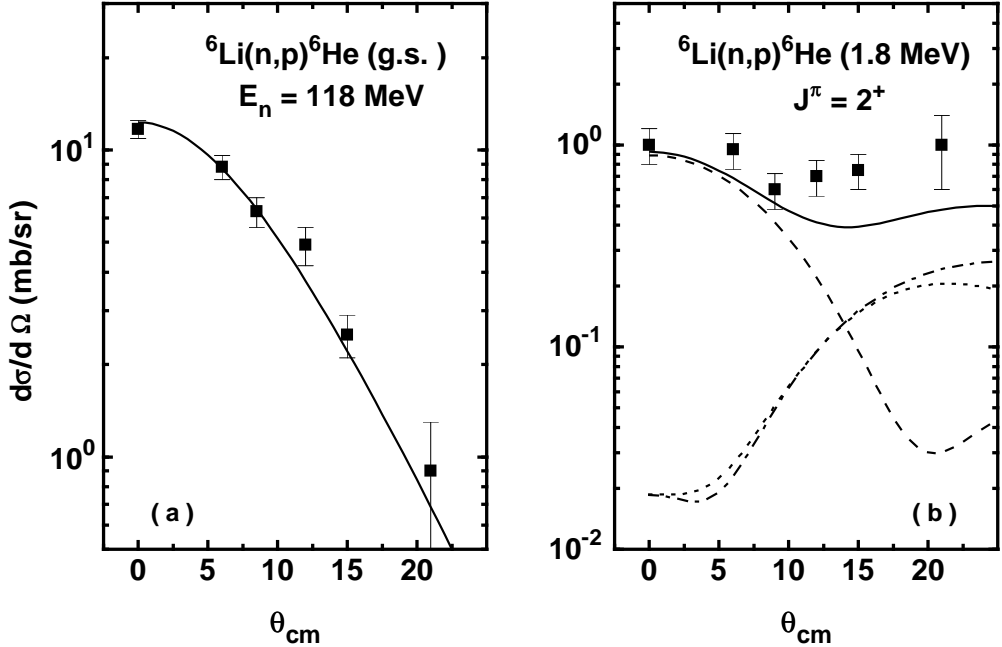}
\vspace*{-10cm}
\caption{Angular distributions for $^6$Li(n,p)$^6$He
at 118~MeV. The dashed, dotted and dashed-dotted lines show the 
contributions from $j=1,2\mbox{ and } 3$, respectively.
The experimental data are from ref.~\protect\cite{wang88}.
}
\label{fig2}
\end{figure}

The second way is to compare our calculation with known experimental data
for excitation to the continuum.
In work \cite{wang88} the reaction $^6$Li(n,p)$^6$He at neutron energy 118 MeV 
was measured.  The proton energy resolution in the experiment was $\sim$ 
2.3 MeV. This should be kept in mind when comparing with the reported
differential cross sections for transitions from the
$1^+$ ground state of $^6$Li to the $0^+$ ground state and $2^+_1$ resonance 
(1.8 MeV) of $^6$He. Fig.~\ref{fig2} shows
the corresponding experimental data plotted together with our calculations using
the Love and Franey $t$-matrix interaction \cite{LoFr} at 100 MeV and an optical
potential \cite{Moake} describing proton elastic scattering from $^6$Li
at 144 MeV. A good description for 
the shape and absolute value of the differential cross section to ground state
was obtained and also a reasonable agreement with the data on the $2^+$ 
resonance. The $2^+$ angular
distribution has a characteristic form corresponding to a transition of
mixed angular momenta. To demonstrate this we show in Fig.~\ref{fig2}b
the separate contributions from transitions with total transferred
$j$ equal 1, 2 and 3 by dashed, dotted and dashed-dotted lines, respectively.
The transition with $j$ = 1 includes transfer of relative orbital momentum 
$l$ = 0 and determines the cross section at small angles, the others with 
$j$ = 2 and 3 have $l$ = 2  and dominate at larger angles. 

Thus the reliability of our approach was confirmed, arguing for our 
predictions for low-energy excitation spectra, for which the model 
was developed.

\subsection  {Inclusive excitation spectra}

\begin{figure}[hbt]
\vspace*{-2cm}
\centering\leavevmode
 \epsfxsize 13.5 cm \epsfbox{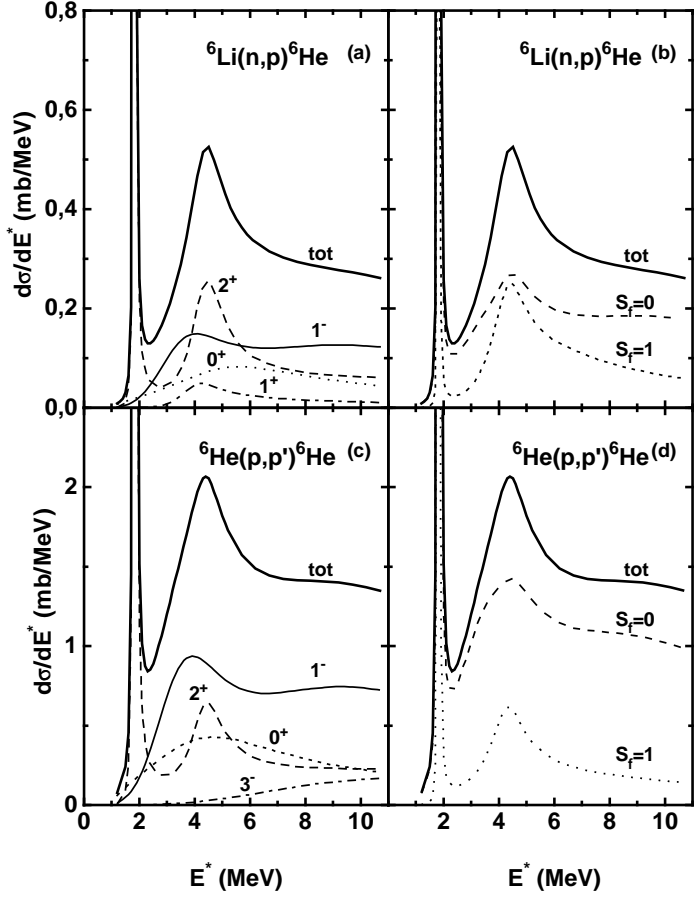}
\vspace*{-3cm}
\caption{Multipole $J^{\pi_f}_f$ and spin decomposition (left and right sides) 
of inclusive proton energy spectra from $^6$Li(n,p)$^6$He$^*$ (top row)
and $^6$He(p,p$'$)$^6$He$^*$ (bottom row) reactions.
}
\label{fig3}
\end{figure}

\begin{figure}[hbt]
\vspace*{0cm}
\centering\leavevmode
 \epsfxsize 12 cm \epsfbox{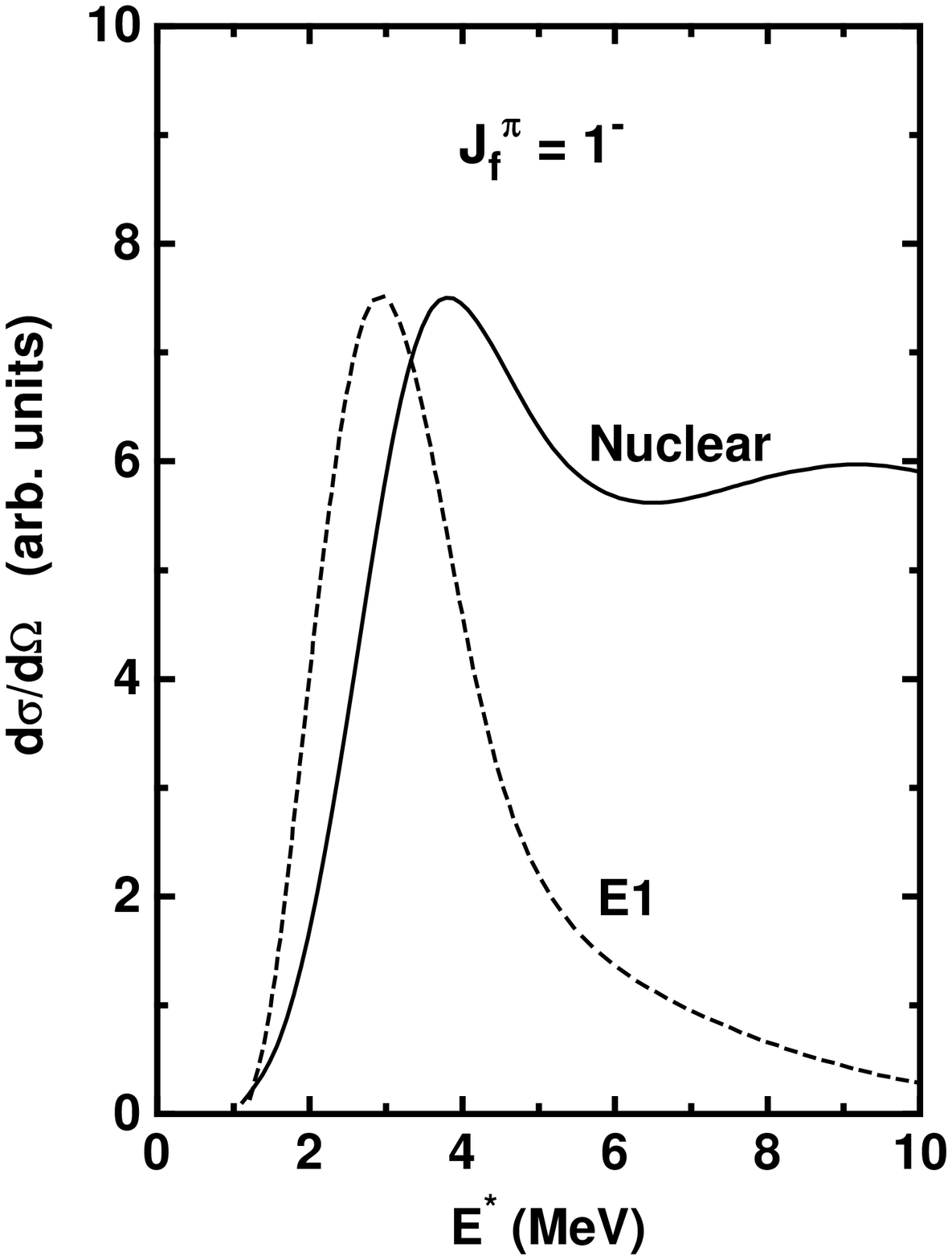}
\vspace*{-2cm}
\caption{The electromagnetic and nuclear inelastic cross section for $1^-$ excitation.}
\label{fig4}
\end{figure}

\subsubsection  {Partial content}

 The inclusive excitation spectra (Fig.~\ref{fig3}, thick solid line) 
for charge-exchange  and inelastic scattering reveal
two distinguished bumps in the low-energy total spectrum : The first,
narrow at excitation energy $\sim$ 1.8 MeV and the second, broad at
$\sim$ 4.5 MeV. To understand the nature of these structures, 
the left side of Fig.~\ref{fig3} shows
the decomposition of the total spectra into contributions from
excitations of different partial components $J_f^{\pi}$ of $^6$He. The $1^-$, 
$2^+$ and $0^+$ excitations are given by thin solid, dashed and dotted lines,
respectively. The dot-dashed line shows the contribution from $1^+$
excitation for charge-exchange and $3^-$ for inelastic scattering.
Contributions from other partial waves are less significant and not given in the
figure.

 The first narrow peak is the well-known $2^+_1$
resonance in $^6$He. The broad bump has a more complex structure. A mixture of
different excitations is responsible for the total shape; a second $2^+$
resonance and concentration of low lying strength of $1^-$  and $0^+$ 
excitations dominate the spectrum. The double-hump shape of $2^+$ excitations is 
the most remarkable feature of the low-energy spectrum. The strength
concentration of $1^-$ transitions at $E_x \sim$ 4 MeV is the other
peculiarity. The behavior of other excitations, for example $3^-$, is different.
It smoothly increases from threshold, and in the case of inelastic scattering 
gives a significant contribution at higher excitation energy.

The excitation spectra for both
reactions have qualitatively the same gross structure, but the absolute
cross sections are a few times larger for inelastic scattering than
for charge-exchange.

Nuclear reactions in which halo nuclei take part serve, due to somewhat 
different 
dynamics, like filters and could lead to different multipole composition in 
observed excitation structures. In inelastic scattering the dipole mode 
dominates while in charge-exchange the $2^+$ resonance is about 
50$\%$ larger. 

The pronounced  $1^-$ nuclear excitation has similarities
with electromagnetic response for the soft dipole mode, prevalent
in Coulomb breakup on heavy targets. Fig.~\ref{fig4} shows theoretical 
cross sections for Coulomb breakup (dotted line) of $^6$He on gold at  
63 MeV/A and inelastic proton scattering (solid line) with $1^-$ 
excitations, arbitrarily normalized. A semiclassical description is used
for the Coulomb dissociation process and our model  
for the electromagnetic dipole response in $^6$He \cite{ferr93}. 
Both processes show strength  accumulation in the same
energy region and hence, we should expect no matter which 
excitation mechanism dominates, a similar behaviour for 
excitation  functions. In an elegant experiment on $^6$He breakup reaction at
63.2 MeV/nucleon on Al and Au targets \cite{balamuth} with registration 
of $\gamma-$rays in coincidence, similar 
behaviour of $\alpha-$particle distributions was found for both targets.
For the light Al target, nuclear mechanism is believed to give the main
contribution to the spectra while for Au the EM dominates. Our theoretical
results explain qualitatively the observed similarity.    

\subsubsection  {Spin structure}

The composition of spectra, or relative
role of excitations of various $J^{\pi_f}_f$, is as discussed above 
different for the two reactions.
In charge-exchange a relatively larger number of states was excited with about 
equal intensity, while inelastic scattering is more selective. 
To better illustrate this point, the right side of Fig.~\ref{fig3} shows
separately the contributions from excitations of $^6$He states with total
spin $S_f$ = 0 (dashed line) and 1 (dotted line). For inelastic scattering
the excitations with $S_f$ = 0 dominate the spectrum, while for charge-exchange
both contributions become comparable. This is a reflection of specific
reaction mechanisms. In inelastic scattering the S = 0, T = 0 component
of effective interactions is the biggest one, while in charge-exchange 
only isovector components play a role and in the charge channel the effective
forces with S = 0 and S = 1 are comparable in strength. The relative role of
different components of effective forces depends on collision energy and
so the ratio between excitations of the various structures will change 
accordingly.

\subsection  {Double-differential inclusive cross sections}

\subsubsection  {Fixed angle}

 Excitation functions, measured at a fixed angle, can serve as a filter for
 selecting partial waves with definite multipolarity and therefore  make it
 possible to extract information on resonances in complex situations
 like that described above. 

\begin{figure}[hbt]
\vspace*{-1cm}
\centering\leavevmode
 \epsfxsize 13.5 cm \epsfbox{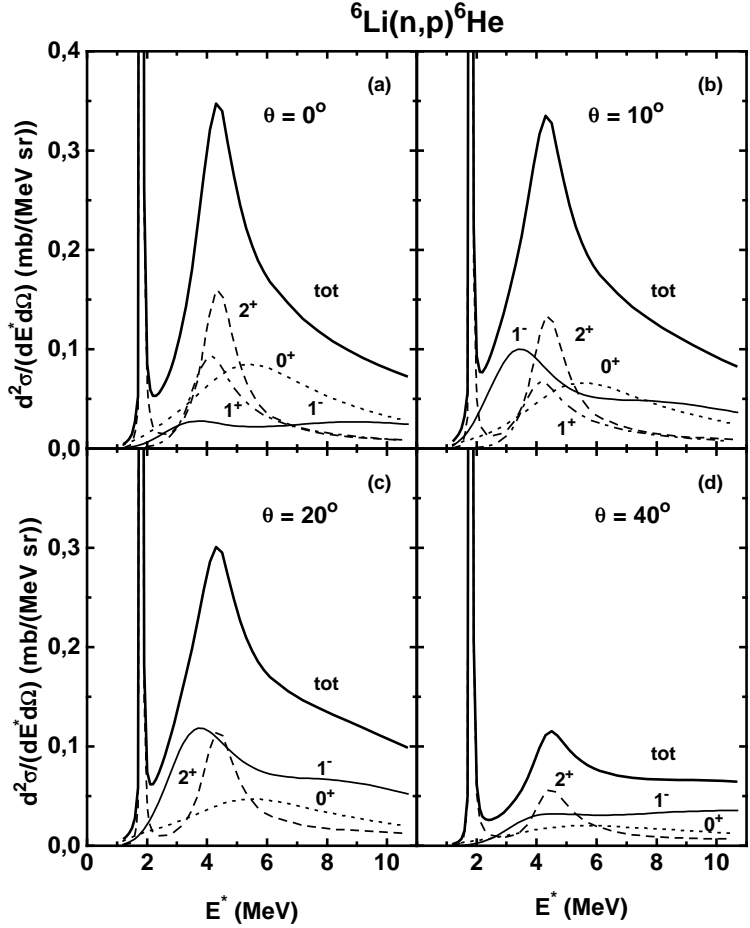}
\vspace*{-4cm}
\caption{Inclusive differential proton energy spectra from $^6$Li(n,p)$^6$He$^*$
for four values of the scattering angle $\theta_{cm}$.
See the text for further details. }
\label{fig5}
\end{figure}

Fig.~\ref{fig5} shows spectra for charge-exchange at different exit
proton angles. The total, $1^-$, $2^+$, $0^+$ and $1^+$ spectra are 
denoted by thick solid, thin solid, dashed, dotted and dashed-dotted lines,
respectively. The double-hump shape appears at all angles, but the composition
of the second bump depends on scattering angle or angular momenta transferred
in the reaction. At 0$^{\circ}$ the excitation of $0^+$, $1^+$ and 
$2^+$  dominates the spectrum. All are populated by strong 
transitions with relative orbital momentum $l$ = 0.  With increasing angle the 
excitation of $1^-$ grows and at 10$^{\circ}$ all 
these states are  important. At 20$^{\circ}$ the $1^-$ and $2^+$
are most pronounced.
At 40$^{\circ}$ the absolute cross section has been reduced to half value, 
and all energy spectra except $2^+$ have flat distributions.
It is also interesting to note that for $1^-$ excitation the main
contribution comes from the transition with $j = 2$,
$s = 1$ and $l = 1$. The dominance of spin-flip transition is due to the
structure of initial and final states where components with $S_i = 1$ and
$S_f = 0$ prevail.

\begin{figure}[hbt]
\vspace*{-1cm}
\centering\leavevmode
   \epsfxsize 13.5 cm \epsfbox{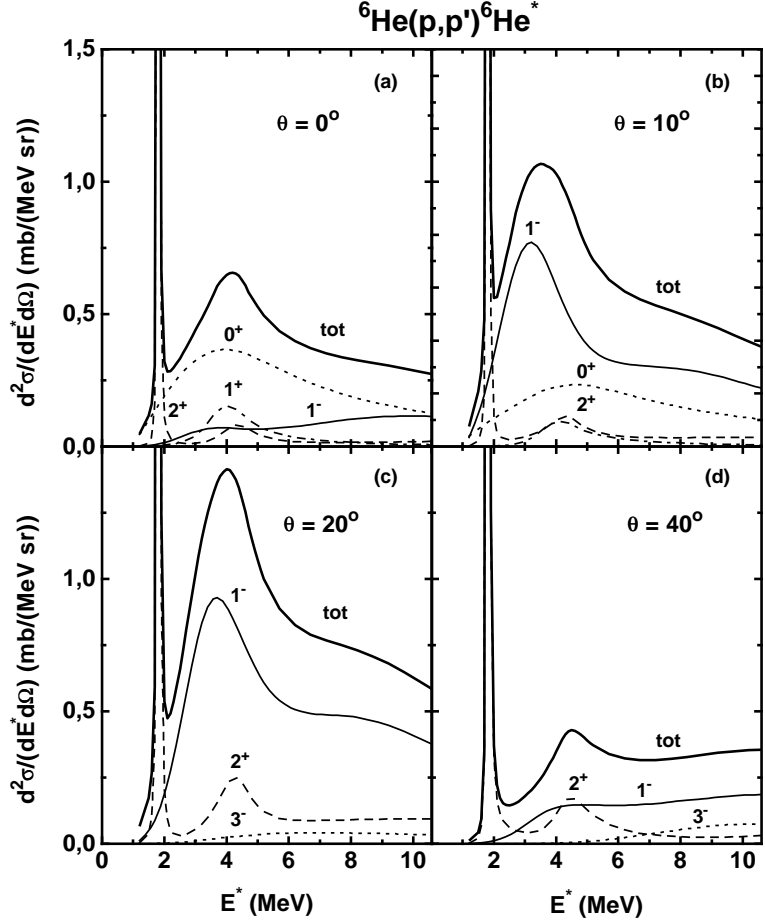}
\vspace*{-4cm}
\caption{Inclusive differential proton energy spectra from $^6$He(p,p$'$)$^6$He$^*$
for four values of the scattering angle $\theta_{cm}$.
See the text for further details. }
\label{fig6}
\end{figure}

Fig.~~\ref{fig6} shows the analogous spectra for inelastic scattering.
All lines mean the same as in the previous figure, except that the dotted line
denotes $3^-$ excitation for $\theta$ = 20$^{\circ}$ and 40$^{\circ}$.
At 0$^{\circ}$ the $0^+$  is excited very effectively in the region of the 
second bump. With increasing scattering angle $1^-$ becomes pronounced. 
It dominates the total
spectra at 10$^{\circ}$ and 20$^{\circ}$. Contributions from $2^+$ and
$1^+$ additionally increase the width of this bump. At 40$^{\circ}$ the total
cross section is again diminished,
with $2^+$ and $1^-$ excitations being the largest. At higher excitation 
energy $3^-$ now gives a significant contribution to the total spectrum.

 From the results represented in the figures, it follows that the second bump
structure in the low-energy part of the spectra is a complex mixture of various
excitations of the $^6$He nucleus. The way these excitations are revealed 
depends on the external fields (or reactions) applied to the system.
This is a characteristic feature of continuum excitations without 
sharp resonances.

\subsubsection  {Fixed excitation energy}

\begin{figure}[hbt]
\vspace*{-1cm}
\centering\leavevmode
 \epsfxsize 13.5 cm \epsfbox{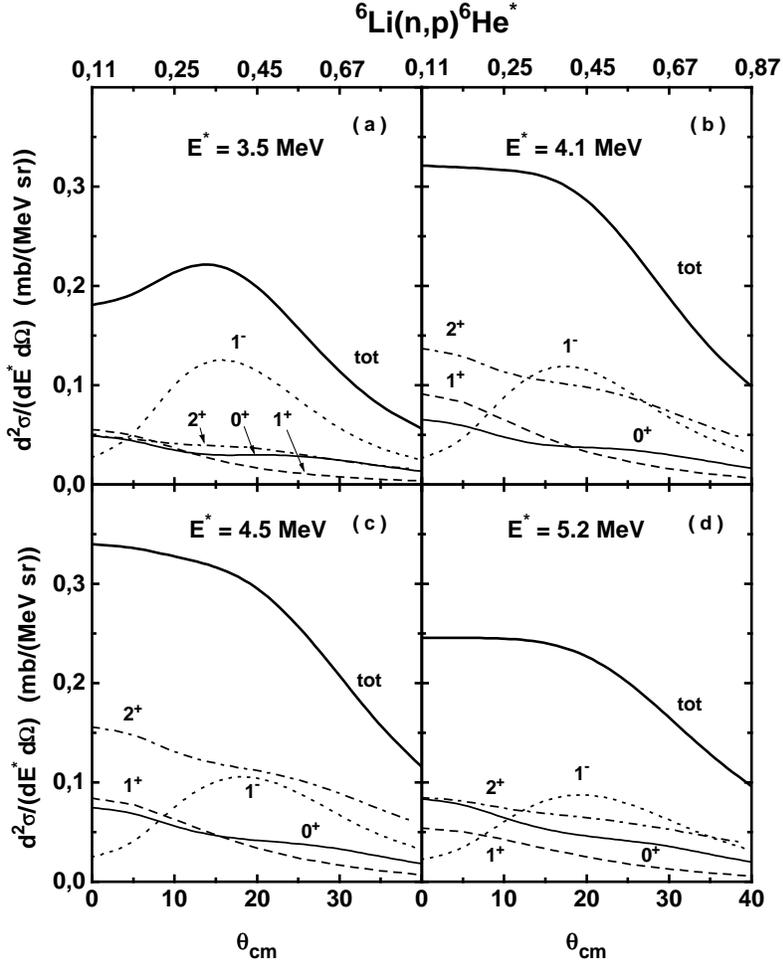}
\vspace*{-4cm}
\caption{Proton energy spectra from the $^6$Li(n,p)$^6$He$^*$
versus scattering angle $\theta_{cm}$ for four
excitation energies $E^*$. 
}
\label{fig7}
\end{figure}

We discussed above the dependence of the differential cross sections on
excitation energy for fixed scattering angle or momentum transfer. It is also
interesting to compare the behavior of the cross sections at fixed excitation
energy for different scattering angles. Fig.~\ref{fig7} shows 
angular distributions for the (n,p) reaction, for a few excitation energies 
that cover both sides of the second bump. The values of the transferred
momenta $\bbox{q}$=$\bbox{k}_i-\bbox{k}_f$ (in units of fm$^{-1}$) are shown
on the top abscissa, corresponding to the scattering angles shown at the bottom
abscissa.
The thick solid line shows the total 
cross section. On the left side of the bump ($E^*$ = 3.5 MeV, Fig.~\ref{fig7}a) 
the differential cross section has an asymmetric bell shape with maximum at 
about 15$^{\circ}$. Going to higher excitation energy ($E^*$ = 4.1 MeV,
Fig.~\ref{fig7}b) through the bump maximum($E^*$ = 4.5 MeV, Fig.~\ref{fig7}c)
to the right side ($E^*$ = 5.2 MeV, Fig.~\ref{fig7}d), the cross section shape 
changes smoothly and becomes
gradually more flat with a plateau from 0$^{\circ}$ to 20$^{\circ}$ on the high
energy side. This shape modification becomes transparent if we examine how the
contributions from excitations of different $J^{\pi _f}_f$ change with energy. 
The thin solid, dotted, dashed and dash-dotted curves show the contributions of
$0^+$, $1^-$, $1^+$ and $2^+$, respectively. We see that at small
angles, the dominating states are  all those ($0^+$, $1^+$ and $2^+$) which can be reached by transitions  
with zero relative orbital momentum. For a range of somewhat larger
angles the contribution of $1^-$ is most significant. The $0^+$ and $1^+$
have smoothly falling angular distributions, $2^+$ is more flat due to the 
already
mentioned mixing of transitions with different $j$. Hence, the interplay
between $0^+$, $1^+$ and $2^+$ excitations, which together
create a smooth background with highest cross section at
small angles, and the $1^-$ peaking at 20$^{\circ}$, define the
total shape. The competition between them is responsible for the modification
of this shape with excitation energy.  As a result, we get a flat total
distribution extending over a rather wide angular range on the high-energy
slope of the second bump. These results are in qualitative agreement with
experimental data on the $^6$Li($^7$Li,$^7Be$)$^6$He$^*$ reaction  
\cite{Sakuta,Janec} if we scale
angular distribution according to the transferred momentum $q$.

\begin{figure}[hbt]
\vspace*{-1cm}
\centering\leavevmode
 \epsfxsize 13.5 cm \epsfbox{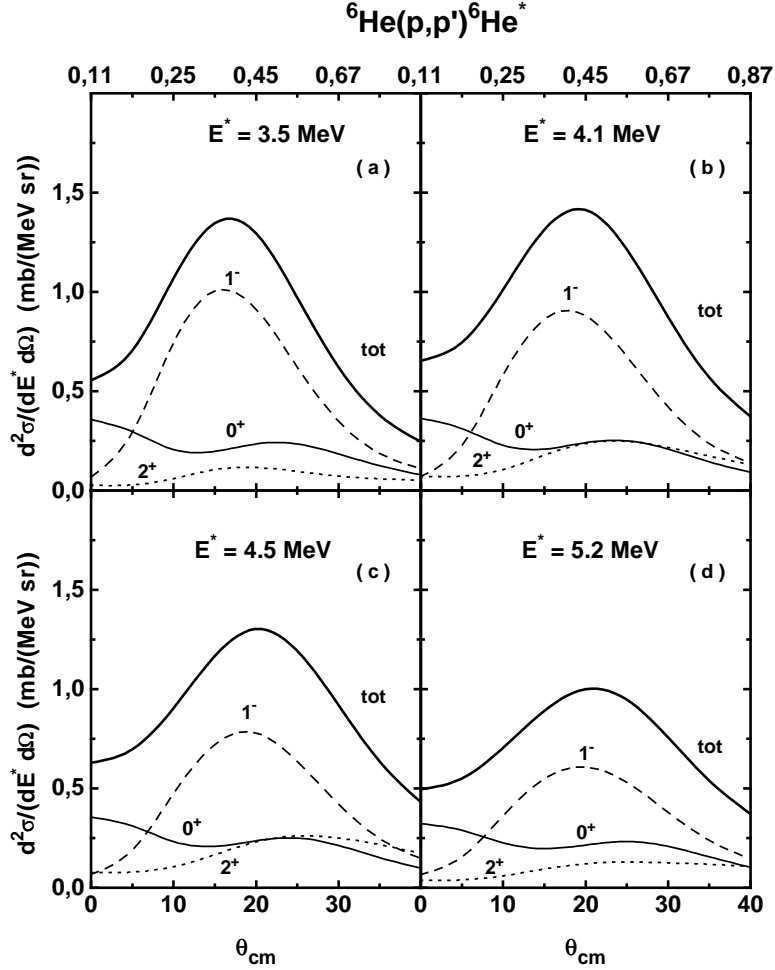}
\vspace*{-4cm}
\caption{Proton energy spectra from $^6$He(p,p$'$)$^6$He$^*$
versus scattering angle $\theta_{cm}$ for four
excitation energies $E^*$.
}
\label{fig8}
\end{figure}

The corresponding data for inelastic scattering are shown in Fig.~\ref{fig8}.
The thick solid, thin solid, dashed and dotted lines again show total, 
$0^+$, $1^-$
and $2^+$ cross sections, respectively. For inelastic scattering, in contrast
to charge-exchange, the total cross section remains bell-shaped at
all excitation energies of the second bump. This is caused by the dipole
excitation which dominates the spectra. The contribution from $0^+$ is also
significant, especially at small scattering angles, counteracting the drop
of the total cross section. The excitation of $2^+$ in (p,p$'$) does
not play the prominent role it does in charge-exchange. Together, $0^+$ and $2^+$,
create a smooth background in angular distributions. 
These
differences in the two reactions are due to two reasons: i) because of different
structure of initial states we need different operators to excite the same
final state in $^6$He and ii) in addition to isovector forces in
charge-exchange the strong isoscalar NN interaction acts between
target and projectile nucleons in inelastic scattering.

\begin{figure}[hbt]
\vspace*{-3cm}
\centering\leavevmode
 \epsfxsize 14 cm \epsfbox{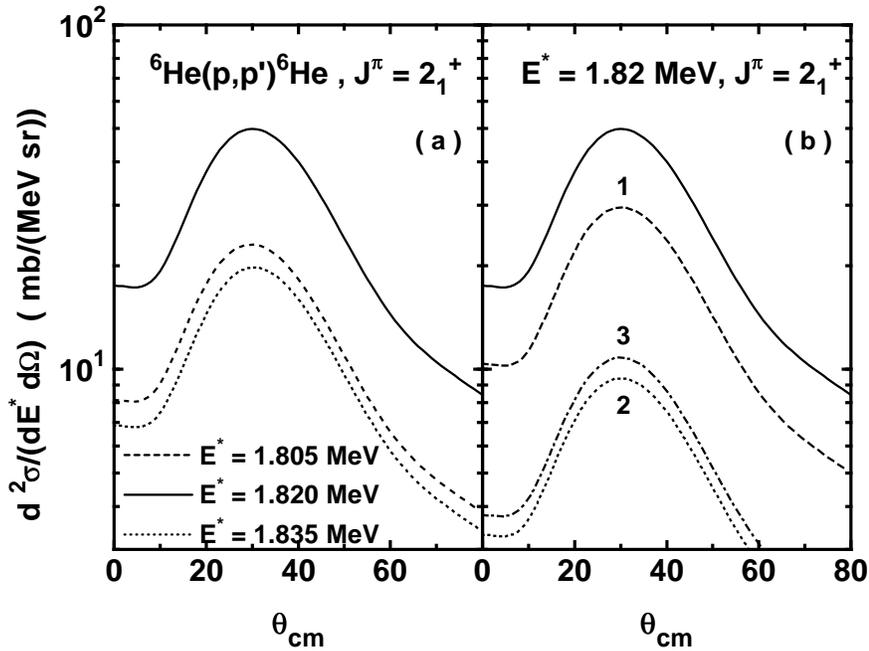}
\vspace*{-6cm}
\caption{Proton energy spectra from $^6$He(p,p$'$)$^6$He$^*$
for the first narrow $2^+$ resonance:
(a) The cross sections at peak position and at energies shifted from the peak
by a half-width to either side. (b) Partial contents of the peak cross section.
For details, see the text.
}
\label{fig9}
\end{figure}

\begin{figure}[hbt]
\vspace*{-1cm}
\centering\leavevmode
 \epsfxsize 13.5 cm \epsfbox{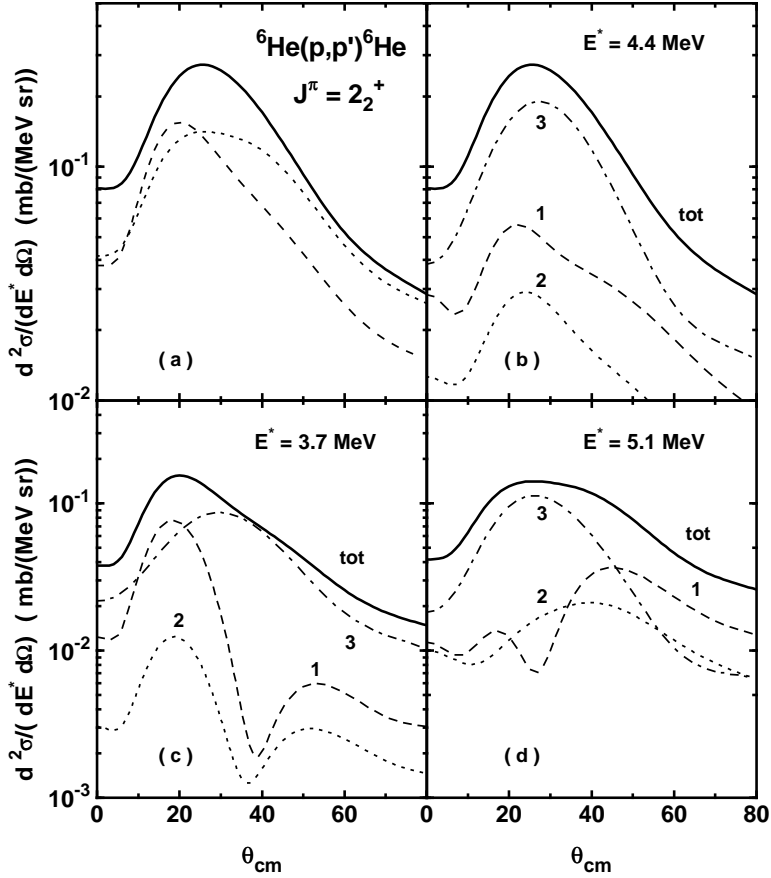}
\vspace*{-4cm}
\caption{Proton energy spectra from $^6$He(p,p$'$)$^6$He$^*$
for the second $2^+$ resonance:
(a) The cross sections at peak position and at energies shifted from the peak
by a half-width to either side. The partial contents of the corresponding cross
sections
are shown in (b), (c) and (d), respectively.  See the text for further details. 
}
\label{fig10}
\end{figure}

To further understand the nature of continuum excitations it is also useful to
make a comparison between angular distributions for the two $2^+$ resonances: 
the first one being narrow
and the second broad. Fig.~\ref{fig9}a shows the differential cross 
sections for $2^+_1$ at three excitation energies: approximately at peak 
(solid line) and at energies shifted
from the peak position a half width to the left (dashed line) and to the
right (dotted line). We see that all angular distributions have identical
shape through the resonance. 
Fig.~\ref{fig9}b shows separately the
contributions from excitation of the three main components of the $^6$He wave
function ( 1 (dashed) is for $L = 2$,
$S = 0$, $l_x = 0$, $l_y = 2$; 2 (dotted) is for $L = 2$, $S = 0$, $l_x = 2$,
$l_y = 0$ and 3 (dot-dashed) is for $L = 1$, $S = 1$, $l_x = 1$, $l_y = 1$)
to the total (solid line) $2^+_1$ cross section at peak position. Again, the
angular distributions for all components have the same shape. For the broad
resonance the picture is different, as shown in Fig.~\ref{fig10}a where 
solid, dashed and dot-dashed
lines  show the total angular distributions for $2^+_2$ at peak
position and shifted from it by a half-width to the left and right, 
respectively.
For the broad resonance, the shape of the differential distribution
is changed through the resonance. 
Fig.~\ref{fig10}b,c,d shows decomposition of the
total distribution into contributions from main components. The notation is 
the same as in Fig.~\ref{fig9}b.
For $2^+_2$ the main contribution comes from excitation of final 
quantum numbers $L = 1$ and $S = 1$ (curve 3). The shape of this
component is only slightly changed  
when going across resonance: the maximum shifts by about 2$^{\circ}$ 
and the width becomes narrower on the high-energy side. The shapes of other 
components experience dramatic changes: the interference pattern in the 
angular distributions
has a different character on opposite sides of the resonance. Usually the
resonance amplitude, as function of energy,  can be separated into a smooth
background and a resonance part. It is reasonable to assume that for a sharp
resonance the background part remains more or less constant over the resonance
width and all energy dependence is
only in the resonance part. As a result, the shape of angular distributions 
does not change over the resonance. For broad resonances the background part 
may change, and interference with the resonance part can produce different
angular distributions. For dominant components the role of the background part 
is relatively small, hence shape variations are not pronounced. For smaller
components both parts of the amplitude are comparable, and can give different
angular distributions on opposite sides of the resonance.

\subsection  {Transition densities to bound and continuum states}

\begin{figure}[hbt]
\vspace*{-1cm}
\centering\leavevmode
 \epsfxsize 11 cm \epsfbox{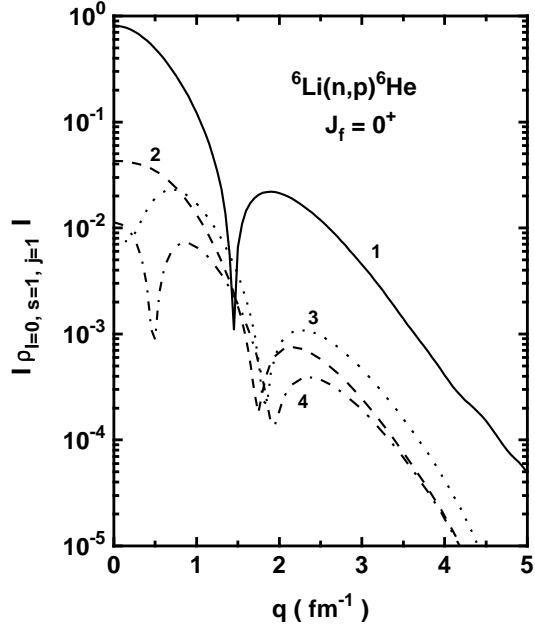}
\vspace*{-4.5cm}
\caption{Absolute values of the charge-exchange momentum space transition 
densities
$\rho_{l=0,s=1,j=1} (q)$ to the $0^+$ states of $^6$He are shown.
Curve 1 is the $0^+_1$ density, curves 2-4 are the $0^+_2$ density components. 
For details, see the text.
}
\label{fig11}
\end{figure}

Another interesting illustration can be obtained from
comparison of transition densities to bound and continuum states. As an example,
Fig.~\ref{fig11} shows transition densities in momentum space from the $1^+$ 
ground state of
$^6$Li to ground and continuum $0^+$ states of $^6$He for
transferred orbital, spin and total angular momenta 
equal 0, 1 and 1, respectively. The continuum energy was chosen
as 6 MeV, where excitation of the continuum
$0^+$ in charge-exchange is largest. Since transition
densities to continuum are complex, we show only absolute values. 
Curve 1 is for transition to ground state, curves 2,3 and 4 are
components of continuum $0^+_2$ with quantum numbers 
$(K=2, L=1, S=1, l_x=1, l_y=1)$, $(K=2, L=0, S=0, l_x=0, l_y=0)$
and $(K=0, L=0, S=0, l_x=0, l_y=0)$, respectively.
We see that between bound states the transition density has a unique spectral
composition, which coincides with that known from the electron scattering M1 
form-factor.
 For transitions to the continuum, there are various components with
different spectral forms. In the bound transition, similar components with the
same quantum numbers also exist, but since the bound state presents a
unique structure all these components are organized in a unique way, and give a
joint system response to external perturbation. In the continuum the various
components correspond to different modes of relative motion between breakup
fragments and should, in principle, be accessible to measurement.  They will be
excited differently by different reactions and the response will depend from 
the external fields applied to the system. Only in the case of sharp resonances
(which in many respects are similar to bound states and represent to a large
extent the internal property of the system) will the response be more or less
the same.

\section{Conclusion}

We have developed a four-body distorted wave theory which is 
appropriate for analysis
of  nucleon-nucleus reactions leading to continuum excitations
of two-neutron Borromean halo nuclei. Spatial granularity of the halo bound 
state and the final state interaction in the 3-body continuum
was fully taken into account by the method of hyperspherical harmonics. 
The weak binding and dilute matter of halo systems enabled us to use a
free NN $t$-matrix for the interaction with halo nucleons. 
Although applicable to any two-neutron Borromean halo nucleus, the
$A=6$ nuclei were again chosen as benchmark systems.
For these nuclei we have the most complete knowledge of the binary subsystems.
Experimental investigations of these nuclei are also currently performed
or planned.
As an initial check the model was successfully tested against data for 
elastic $^6$Li(p,p)$^6$Li, inelastic $^6$Li(p,p$'$)$^6$Li ($0^+$, 3.56 MeV) and 
$^6$Li(n,p)$^6$He (gs.\ and $2^+_1$, 1.8~MeV), which has been available
for a decade.

A detailed study
of inclusive excitation and differential cross sections for inelastic
$^6$He(p,p$'$)$^6$He$^*$ and charge-exchange $^6$Li(n,p)$^6$He$^*$ reactions at
beam energy 50 MeV was performed. The theoretical low-energy
spectra exhibit two resonance-like structures. The first (narrow) 
is the excitation of the well-known $2^+_1$ resonance. The second (broad) bump
is a structural composition of overlapping soft modes of multipolarities 
$1^-, 2^+, 1^+, 0^+$ whose relative weights depend on transferred
momentum and reaction type. Recent experimental data on heavy-ion
charge-exchange reactions \cite{Sakuta,Janec}, although sparse,
confirm the existence of the second structure.

The soft excitations of different multipolarities have
a concentration in a relatively narrow energy region near the $2^+_1$ 
resonance. This poses a challenge.
Nuclear reactions in which halo nuclei take part serve however, due to differences 
in reaction mechanisms, like filters emphasizing different multipole 
components in the observed excitation structures. To some extent we may exploit
this to our advantage.

Thus comparison of (n,p) and (p,p$'$) shows that the excitation cross section for
inelastic scattering preferentially selects the $1^-$ component. Hence (p,p$'$)
is the most promising tool for studying the soft dipole excitation 
mode.

Double differential distributions for the broad structure show that 
association of the observed structure with excitation of a unique
multipolarity would be misleading. This is especially so for
charge-exchange $^6$Li(n,p)$^6$He$^*$, where a flat shape 
of the total angular distribution extending outside forward angles, is due to 
mixing of excitations with different multipolarities.
Under favorable conditions, measurement of 
spectra at definite momentum transfer makes it possible to extract 
information on individual resonances in complex situations like the one 
described above.

Our results on charge-exchange are in qualitative agreement with
experimental data on the $^6$Li($^7$Li,$^7$Be)$^6$He$^*$ reaction  
\cite{Sakuta,Janec} if we scale
angular distributions according to the transferred momentum. Forward angles 
are most important for partial analysis, but in
both experiments there is not enough statistics in this region for more 
definite conclusions on the resonant structure of $^6$He continuum.
Since all resonant states are concentrated in the vicinity of the extremely
pronounced $2^+_1$ state, high resolution experiments
with detailed angular distributions will be needed.

The model we have developed allows us to calculate cross sections for kinematically 
complete experiments when characteristics of four particles are measured.
Hence we can study a variety of correlations existing in Borromean halo
nuclei that could not be seen in the inclusive observables. 
An analysis of different exclusive cross sections of nucleon-nucleus
reactions with excitation of the 3-body continuum of $^6$He is in progress.

\section{Acknowledgements}
This work was done under financial support from the Bergen and Surrey
members of the RNBT collaboration. Two of the authors (B.D.\ and S.E.) are
thankful to the University of Bergen and University of Surrey for hospitality.
Two of the authors (S.E.\ and F.G.) acknowledge support from RFFI grant
96-02-17216. Support from ECT$^*$, Trento, Italy (J.S.V.) and NORDITA, 
Copenhagen (J.S.V., B.D.) where some of the work was carried out,
is furthermore acknowledged. The authors are 
grateful to Prof.\ M.V. Zhukov for useful discussions.


\appendix
\section{}

Within the cluster representation (for details see Refs.
\cite{prep93,dancon,danden}), 3-body bound and continuum state wave
functions (WF) have the product form

\begin{equation}
\mid \lwf{\Phi} >\ = \exp(i\bbox{K}\cdot\bbox{R})\
\lwf{\Phi} _{C}(\zeta _{C})  \lwf{\Psi} ^{T}_{JM}
\end{equation}

\noindent
where  $\Phi _{C}(\zeta _{C})$ is an intrinsic core WF, while
 $\Psi ^{T}_{JM}$ is the ``active'' part of the
3-body  WF carrying the total angular moment $J$, its projection $M$
and total isospin $T$. This part depends on relative coordinates and cluster
spins (suppressed in our notations) and it is the object of the calculation.
$\bbox{K}$ and $\bbox{R}$ are momentum and  coordinate  of the center of mass of the
nucleus $A$,
respectively,

Translationally invariant normalized sets of Jacobi coordinates $\bbox{x}_3$ 
and $\bbox{y}_3$ are defined as follows

\begin{eqnarray}
\bbox{x}_3 &=& \sqrt{ A_{12} }\ (\bbox{r}_{2}-\bbox{r}_{1}),\nonumber \\
\bbox{y}_3 &=& \sqrt{ A_{(12)C} }\
( \bbox{r}_{C}- {A_{1}\bbox{r}_{1}+A_{2}\bbox{r}_{2} \over A_{1}+A_{2} } ), \\
\bbox{R} &=& {1 \over A}\ (A_1\bbox{r}_1+A_2\bbox{r}_2+A_C\bbox{r}_C).  \nonumber
\end{eqnarray}

\noindent
Here $A_{12}=A_{1}A_{2}/(A_{1}+A_{2})$ is the reduced mass
of the (12) subsystem in units of the nucleon mass $m_N$,
$A_{(12)C}=(A_{1}+A_{2})A_{C}/(A_{1}+A_{2}+A_{C})$ is the reduced mass
of the (12) cluster with respect to the core C, and
$A=A_1+A_2+A_C$. Notice, that $\bbox{y}_{3}$ is co-linear with
$\bbox{{r}_C-{R}}$. Alternative sets $(\bbox{{x}_1,{y}_1})$ and
$(\bbox{{x}_2,{y}_2})$ of Jacobi coordinates are obtained by cyclic
permutations of (1,2,C).
The set of Jacobi momenta $\bbox{q}_3$, $\bbox{p}_3$ and $\bbox{K}$ conjugate
to $\bbox{x}_3,\bbox{y}_3 \mbox{ and }\bbox{R}$ is defined by the relations,

\begin{eqnarray}
\bbox{q}_3 & = & \sqrt{ A_{12} }\ ({ \bbox{k}_{1} \over A_1} -
{ \bbox{k}_{2} \over A_2} ), \nonumber \\
\bbox{p}_3 & = & \sqrt{ A_{(12)C} }\
 ({ \bbox{k}_{C} \over A_C} - { \bbox{k}_{1}+\bbox{k}_{2} \over A_1+A_2 } ), \\
\bbox{K} & = & \bbox{k}_{1}+\bbox{k}_{2}+\bbox{k}_{C}, \nonumber
\end{eqnarray}

\noindent
where $\bbox{k}_i,\,\,\, i=1,2,C$ are the particle wave numbers in an
arbitrary frame. The Jacobi momenta $\bbox{q}_3$, $\bbox{p}_3$ are connected to
$\bbox{k}_x$ and $\bbox{k}_y$ defined in (\ref{eq:defk}) with simple relation
\begin{eqnarray}
\bbox{k}_{x} & = & \sqrt{ A_{12} }\ \bbox{q}_3, \nonumber \\
\bbox{k}_{y} & = & \sqrt{ A_{(12)C} }\ \bbox{p}_3,
\end{eqnarray}

\noindent
We use hyperspherical coordinates $\rho$, $\alpha$, $\theta_x$, $\phi_x$,
$\theta_y$, $\phi_y$, where ($\theta_x$, $\phi_x$) and ($\theta_y$, $\phi_y$)
are angles associated with the unit vectors $\bbox{\hat{x}}$ and $\bbox{\hat{y}}$,
and

\begin{equation}
\rho  = (x^{2}+ y^{2})^{1/2}=(\sum_{i=1,2,C} A_i ( \bbox{
r_i-R})^2\ )^{1/2},\,\,\,\,\,\,\,\,\,\,\,\,\,\,\,\,\,\,\,
\alpha=\arctan(x/y).
\end{equation}

\noindent
The collective variables $\alpha$  and $\rho$ are called
hyperangle and hyperradius. The last variable is  rotationally and
permutationally invariant, having the character of total moment of
inertia or a weighted measure of distances in the 3-body system. The corresponding
conjugated momenta are

\begin{eqnarray}
\kappa  = (q^2+ p^2)^{1/2}= \hbar ^{-1}(2m_N\mid E_{\kappa}\mid)^{1/2} 
\mbox{ ,   } \alpha_{\kappa}  = \arctan(q/p)
\end{eqnarray}

\noindent
where $E_{\kappa}$ is the total 3-body energy.
Since we introduce a new degree of freedom its corresponding conjugated
quantum operator has eigenvalues $K=2n+l_x+l_y$ called the hypermoments.
Hyperspherical harmonics (HH)
$\lwf{\psi}_K^{l_xl_y}(\alpha)\cdot Y_{l_xm_x}(\Omega_x)
\cdot Y_{l_ym_y}(\Omega_y)$
are eigenfunctions of of this operator.

We seek  our bound-state  and continuum wave functions in the form of
expansions on a generalized angle-spin basis ($LS$ coupling)

\begin{equation}
\Upsilon^{l_{x}l_y}_{JKLSM_J}(\Omega_{5}) = \Bigl[ { {\cal Y}}^{l_{x}l_y}_{KL}
(\Omega_{5}) \otimes X_{S}\Bigr]_{JM_J}
\end{equation}

\noindent
with HH defined as

\begin{equation}
{\cal Y}^{l_xl_y}_{KLM}(\Omega_5)=\lwf{\psi}_K^{l_xl_y}(\alpha)\Bigl[
Y_{l_x}(\Omega_x) \otimes Y_{l_y}(\Omega_y) \Bigr]_{LM}
\end{equation}

\noindent
Here the $\alpha$, $\theta_x$, $\phi_x$, $\theta_y$ and $\phi_y$ variables are
denoted collectively by $\Omega_5$, $X_{S}$ is a spin function,
$\Bigl[\ldots \otimes \ldots \Bigr]$ means vector coupling,

\begin{equation}
\Bigl[ A_l \otimes B_{\lambda} \Bigr]_{jm} = \sum_{m_l, m_{\lambda}}
<l m_l\ \lambda m_{\lambda} \mid j m >\ A_{lm_l} \ B_{\lambda m_{\lambda}}
\end{equation}

\noindent
The relative orbital momenta $l_x$, $l_y$, couple to the total
orbital momentum $L$ and its projection $M$. Hyperangular part of HH has the
following explicit form

\begin{equation}
\lwf{\psi}^{l_{x}l_{y}}_{K}{(\alpha )}\ = N^{l_x l_y}_{K}\
(\sin\alpha)^{l_x}\ (\cos\alpha )^{l_y}\ {\displaystyle
P_{{ K-l_x-l_y \over 2}}^{l_x+1/2,l_y+1/2}(\cos2\alpha) },
\end{equation}

\noindent
where $P_{n}^{\alpha,\beta}$ are Jacobi polynomials and $N^{l_x l_y}_{K}$ is
a normalization factor.

For bound states the internal WF in LS coupling has the form

\begin{equation}
 \lwf{\Psi}^{T}_{JM} \ =
{\displaystyle
{1 \over \rho^{5/2}}\ \sum_{\gamma}\ \lwf{\chi}_{Kl_xl_y}^{LS}(\rho)\
\Upsilon^{l_{x}l_y}_{JKLSM_J}(\Omega_{5})\ X_{TM_T} }.   \label{eq:expan}
\end{equation}

\noindent
where $\gamma$ is an abbreviation for a set of quantum numbers
$\gamma = \{K, L, S, l_x, l_y \}$. For continuum states we have the
following form

\begin{eqnarray}
 \lwf{\Psi}^{T}_{JM}  &=&
{\displaystyle
{1 \over (\kappa\rho)^{5/2}}\ \sum_{\gamma, \gamma'}\
\lwf{\chi}_{Kl_xl_y,K'l'_xl'_y}^{LS,L'S'}(\kappa\rho)\
\Upsilon^{l_{x}l_y}_{JKLSM_J}(\Omega_{5}) }
\nonumber \\
&\times&
<L'M'_L S'M'_S \mid JM>\ {\cal Y}^{l'_{x}l'_y \ \mbox{\large $*$}}_{K'L'M'_{L}}
(\Omega_{5}^{\kappa})\  X_{TM_T}
\end{eqnarray}

\noindent
with normalization condition

\begin{equation}
\int  \lwf{\Psi}_{\kappa'}^{\mbox{\large $*$}}\lwf{\Psi}_{\kappa}d\bbox{x}d\bbox{y}= \kappa^{-5}
\delta (\kappa'-\kappa)\delta (\Omega_5^{\kappa'}-\Omega_5^{\kappa})
=\delta (\bbox{q'}-\bbox{q})\delta (\bbox{p'}-\bbox{p})
\end{equation}

\noindent
The WF $\Psi_{JM}^T$ is a solution of the 3-body Schr\"odinger equation

\begin{equation}
(\hat{T} +\hat{V} -E) \lwf{\Psi}_{JM}^T=0 \mbox{ ,    } \hat V
=\hat V_{12}+\hat V_{1C}+\hat V_{2C},
\end{equation}

\noindent
where $\hat{V}_{ij}$ is the interaction potential between particles  i and j.
After separating out the hyperangular parts of the WF we obtain a set of
coupled equations similar  to those for a  particle moving in a  deformed
mean field.

In case of neutral particles the bound hyperradial WF for Borromean nuclei
has a true 3-body asymptotics

\begin{equation}
\lwf{\chi}_{\gamma}(\rho \rightarrow 0) \sim \rho^{K+5/2} \mbox{ ; }
\qquad \lwf{\chi}_{\gamma }(\rho \rightarrow \infty ) \sim \exp(-\kappa \rho) 
\end{equation}

\noindent
For continuum WF the boundary condition at the origin  coincides
with that for the bound state, while for chargeless  particles at
$\rho \rightarrow \infty $ it is

\begin{equation}
\lwf{\chi}_{\gamma , \gamma'}(\kappa \rho ) \sim
\rho^{1/2} (H^{-}_{K+2}(\kappa \rho)\delta_{\gamma , \gamma'}
- S_{\gamma , \gamma'} H^{+}_{K'+2}(\kappa \rho ))
\end{equation}

\noindent
Here $H^{-}_n$and $H^{+}_n$ are Hankel functions of integer index $(n =K+2)$
with asymptotic
$\sim {\displaystyle {1 \over \sqrt{\rho}}} \exp(\mp i\kappa \rho )$,
describing the in-  and  outgoing  3-body spherical waves,
$S_{\gamma ,\gamma'}$ is the $S$-matrix for the $3\rightarrow 3$  scattering.

Wave functions discussed above are characterized by the total angular momentum
$J$ and its projection $M$. Due to rotational invariance the continuum wave
functions with different $J$ are dynamically decoupled and can be calculated
separately. For transition densities we need the 3-particle scattering
states $\Psi_{m_1, m_2}^{(+)} (\bbox{k}_x, \bbox{k}_y)$ in other
representation  characterized by $\bbox{k}_x$ and $\bbox{k}_y$ momenta of
relative motions and projections $m_1$ and $m_2$ of particle spins on a
chosen direction. They can be written as follows

\begin{eqnarray}
 \lwf{\Psi}_{m_1, m_2}^{(+)}  &=& {\displaystyle
{1 \over (\kappa\rho)^{5/2}}\ \sum_{\gamma, \gamma',J, M, M'_L}
\lwf{\chi}_{Kl_xl_y,K'l'_xl'_y}^{LS,L'S'}(\kappa\rho)\
\Upsilon^{l_{x}l_y}_{JKLSM}(\Omega_{5})\
{\cal Y}^{l'_{x}l'_y \ \mbox{\large $*$}}_{K' L' M'_{L}}(\Omega_{5}^{\kappa}) } \nonumber \\
&\times& <L'M'_L S'M'_S \mid JM>\
<s_1 m_1\ s_2 m_2 \mid S' M_S'>\ X_{TM_T}.
\end{eqnarray}

\noindent
The transition density describes the system response to a zero-range
perturbation, and can be expressed as a matrix element between initial bound
and final continuum states

\begin{eqnarray}
\rho_{mM_T}^{lsj,T} &=& < \lwf{\Psi}^{(-)}_{m_1, m_2} \mid \sum_{t = 1,2}
{\delta(r-r_t) \over r_t^2 } \Bigl[ Y_l(\bbox{\hat{r}}_t) \otimes \sigma_t^s
\Bigr]_{jm} \tau_{M_T}^T(t) \mid J_i M_i >  \nonumber \\
&=&  \sum_{J_f, M_f} {1 \over \hat{J}_f } <J_i M_i \ j m \mid J_f M_f >
\sum_{\gamma_f', M_{L_f}', M_{S_f}'} <s_1 m_1 \ s_2 m_2 \mid
S_f' M_{S_f}' > \nonumber \\
&\times& <L_f' M_{L_f}' \ S_f' M_{S_f}' \mid J_f M_f > \
 {\cal Y}_{K_f' L_f' M_{L_f}'}^{l_{x_f}' l_{y_f}' \mbox{\large $*$}}
 (\Omega_5^{\kappa}) \
\rho_{\gamma_f'}^{l s j, T}(r,\kappa) \label{eq:trden}
\end{eqnarray}
 
\noindent
The easiest way to calculate space integrals in $\rho_{\gamma_f'}^{l s j,
T}(r,\kappa)$ is to do it
in a coordinate system where radius $\bbox {r_t}$ is colinear with the $\bbox{y}$-coordinate:

\begin{equation}
\bbox {r_t} = a_k \, \bbox{y_k} = \sqrt{\frac{A_{i} + A_{j}}{A A_{k}}} \:
\bbox{y_k}
\end{equation}

In our case we must rotate from the initial Jacobi coordinate system (basis
$\bbox{x_3, y_3}$ with $A_1=A_2=1$) to the
alternative similar sets $(\bbox{x_1, y_1})$ or $(\bbox{x_2, y_2})$. 
Hyperharmonics ${\cal Y}^{l_x,l_y}_{K L M}$ transform under this rotation 
through Raynal-Revai coefficients

\begin{equation}
{\cal Y}^{l_x,l_y}_{K L M} (\Omega_5) = \sum_{l'_x, l'_y} <l'_x, l'_y | l_x, l_y>_{K L}
{\cal Y}^{l'_x,l'_y}_{K L M} (\Omega'_5)
\end{equation}

\noindent
Using the following definition of reduced matrix elements

\begin{equation}
<j_f m_f | \hat{O}_{j m} | J_i m_i > = { <j_i m_i \ j m |\ j_f m_f > \over
\hat{j}_f } < j_f || \hat{O}_j || j_i>
\end{equation}

\noindent
and with the necessary summation over Clebsh-Gordon coefficients,
the radial part of transition density matrix elements 
$\rho_{\gamma_f'}^{l s j, T}(r,\kappa)$ is
 
\begin{eqnarray}
\rho_{\gamma_f'}^{lsj,T}(r,\kappa) &=& \sum_{\gamma_f, \gamma_i,
l_{x_f}'', l_{y_f}'', l_{x_i}'', l_{y_i}''}  <l_{x_f}'' l_{y_f}'' | l_{x_f}
l_{y_f}>_{K_f L_f} <l_{x_i}'' l_{y_i}'' | l_{x_i} l_{y_i}>_{K_i L_i} \nonumber \\
&\times& < S_f || \sigma^s (1) || S_i> <l_{y_f}'' || Y_l || l_{y_i}''>
\ \hat{j} \hat{L}_i \hat{L}_f \hat{J}_i \hat{J}_f\
\delta_{l_{x_f}'',l_{x_i}''} \nonumber \\
&\times& (-1)^{l_{x_i}''+l_{y_i}''+l+L_f}
{ \left\{ \begin{array}{ccc}
l_{y_i}'' & l_{x_i}'' & L_i       \\
L_f       & l         & l_{y_f}'' \\
\end{array} \right\} }
{ \left\{ \begin{array}{ccc}
S_i & S_f & s  \\
L_i & L_f & l  \\
J_i & J_f & j  \\
\end{array} \right\} } \\
&\times& (1 + (-1)^{S_i+S_f+T_i+T_f})\ \lwf{I}_{\gamma_f',\gamma_f,\gamma_i}
(r, \kappa)\ < T_f M_{T_f} | \tau^T_{M_T} (1) | T_i M_{T_i}> \nonumber
\end{eqnarray}

\noindent
The factor $(-1)^{S_i+S_f+T_i+T_f}$ comes from symmetry properties of spin
and isospin matrix elements
\begin{equation}
<S_f M_f | \sigma^s_{m} (2) | S_i M_i> = (-1)^{S_i+S_f}
<S_f M_f | \sigma^s_{m} (1) | S_i M_i>,
\end{equation}
and the reduced spin and orbital matrix elements are

\begin{eqnarray}
< S_f || \sigma^s (1) || S_i> &=& (-1)^{1+s+S_i} \sqrt{2}\ \hat{s}\ \hat{S}_i\
\hat{S}_f\
{ \left\{ \begin{array}{ccc}
{1 \over 2} & {1 \over 2} & S_i         \\
S_f         & s           & {1 \over 2} \\
\end{array} \right\} } \\
<l_{y_f}'' || Y_l || l_{y_i}''> &=& {1 \over \sqrt{4 \pi} }\ \hat{l}\
\hat{l}_{y_i}''\ (l_{y_i}'' 0\ l 0 \mid l_{y_f}'') \\
\end{eqnarray}

\noindent
The radial matrix element $I_{\gamma_f',\gamma_f,\gamma_i}$ is

\begin{eqnarray}
\lwf{I}_{\gamma_f',\gamma_f,\gamma_i} (r, \kappa) &=& \int_0^{\infty} dx\ x^2\
\int_0^{\infty} dy\ y^2\
{\displaystyle {1 \over (\kappa \rho)^{{5 \over 2}} } }
\lwf{\chi}_{K_f l_{x_f} l_{y_f}, K_f' l_{x_f}' l_{y_f}' }^{L_f S_f, L_f' S_f' \ \mbox{\large $*$}}
(\kappa\rho) \nonumber \\
&\times& {\displaystyle
{\delta(r-ay) \over (ay)^2} \lwf{\chi}_{K_i l_{x_i} l_{y_i}}^{L_i S_i} (\rho)
{1 \over \rho^{{5\over 2}} }\ \lwf{\psi}^{l_{x_f}'' l_{y_f}''}_{K_f}(\alpha)\
\lwf{\psi}^{l_{x_i}'' l_{y_i}''}_{K_i} (\alpha) },
\end{eqnarray}

\noindent
which can be reduced to a one-dimensional integral over the $\rho$-variable

\begin{eqnarray}
\lwf{I}_{\gamma_f',\gamma_f,\gamma_i} (r, \kappa) &=& {\displaystyle
{1 \over a^3 \kappa^{{5 \over 2}}}\int_{{r \over a}}^{\infty} d\rho
{\sqrt{\rho^2 - ({\displaystyle{r \over a})^2}} \over \rho^4  }
\ \lwf{\psi}^{l_{x_f}'' l_{y_f}''}_{K_f}(\alpha)\ \lwf{\psi}^{l_{x_i}'' l_{y_i}''}_{K_i}
(\alpha) }\nonumber \\
&\times& \lwf{\chi}_{K_f l_{x_f} l_{y_f}, K_f' l_{x_f}' l_{y_f}' }^{L_f S_f, L_f'
S_f'\ \mbox{\large $*$}} (\kappa\rho)\ \lwf{\chi}_{K_i l_{x_i} l_{y_i}}^{L_i S_i} (\rho)
\end{eqnarray}

\noindent
where $\cos\alpha = {\displaystyle {r \over a\rho} }$ and
$a = a_1 = a_2$.

\section{}

To calculate the radial formfactors we need the multipole decomposition
of the effective NN interaction. For this purpose  it is 
convenient to use momentum representation

\begin{equation}
 V(\bbox{r}_{pt},\bbox{p}_{pt} ) = {1\over (2\pi)^3}
\int d\bbox{k}\ \exp(-i\bbox{k} \cdot \bbox{r}_{pt})\ V(\bbox{k},\bbox{p}_{pt}) 
\end{equation}

\noindent
where in $V(\bbox{k},\bbox{p}_{pt} )$ the longitudinal 
$t_{T}^{\parallel}(k) = t_{1T}^{C}(k)-2t_{T}^{T}(k)$
and transverse $t_{T}^{\bot}(k) = t_{1T}^{C}(k)+t_{T}^{T}(k)$ parts are
usually singled out

\begin{eqnarray}
V(\bbox{k}, \bbox{p}_{pt}) &=&\sum_{T} { \left\{ \sum_{S}
t_{ST}^{C}(k)\sigma_{p}^{S}\cdot \sigma_{t}^{S} + {i \over k^{2}}
t_{LS}^{T}(k)\bbox{k}\times\bbox{p}_{pt}\cdot\bbox{S}-
t_{T}^{T}(k)S_{pt}(\bbox{\hat{k}}) \right\} } \tau_{p}^{T}\cdot\tau_{t}^{T} \nonumber\\
 &=& \sum_{T}  \Bigl\{ t_{0T}^{C}(k)+t_{T}^{\parallel}(k)
(\bbox{\sigma}_{p}\cdot\bbox{\hat{k}})(\bbox{\sigma}_{t}\cdot\bbox{\hat{k}})+
t_{T}^{\bot}(k)[\bbox{\sigma}_{p}\times\bbox{\hat{k}}]\cdot
[\bbox{\sigma}_{t}\times\bbox{\hat{k}}]  \nonumber \\
 & &+{i \over k^{2}} t_{LS}^{T}(k)\bbox{k}\times\bbox{p}_{pt}\cdot\bbox{S}
 \Bigr\} \tau_{p}^{T}\cdot\tau_{t}^{T}
\end{eqnarray}

\noindent
Formfactors $t^i_j (k)$ are Fourier transforms of corresponding forces in 
coordinate space

\begin{eqnarray}
t_{ST}^{C}(k) &=& 4\pi \int_{0}^{\infty} j_{0}(kr)t_{ST}^{C}(r)r^{2}dr , \\ 
t_{T}^{T}(k)  &=& 4\pi \int_{0}^{\infty} j_{2}(kr)t_{T}^{T}(r)r^{2}dr , \\
t_{LS}^{T}(k) &=& 4\pi k \int_{0}^{\infty} j_{1}(kr)t_{LS}^{T}(r)r^{3}dr .
\end{eqnarray}

\noindent
With shorthand notations for multipole operators,
\begin{eqnarray}
\hat{\rho}_{lsj,m} (i) & = & j_l (kr_i) [Y_l (\bbox{\hat{r}}_i) 
 \otimes \sigma_i^s ]_{jm} \\
\hat{\rho}^l_{lj,m} (i) & = & {1 \over k} j_l (kr_i) [Y_l (\bbox{\hat{r}}_i) \otimes 
 \bbox{\nabla}_i ]_{jm} \\
\hat{\rho}^{ls}_{j,m} (i) & = & {1 \over k^2} (\bbox{\nabla}_i \hat{\rho}_{j0j,m} (i) )
 \times \bbox{p}_i \cdot \bbox{\sigma}_i
\end{eqnarray}
the multipole decomposition of the NN
potential can be written as follows \cite{Petrdec}

\begin{eqnarray}
V(\bbox{r}_{pt}, \bbox{p}_{pt}) &=&\sum_{jT} \tau_{p}^{T}\cdot\tau_{t}^{T}\ 
{2\over\pi}\int_0^{\infty} dk\ k^2\left\{\ t_{0T}(k)\ (\hat{\rho}_{j0j} (p) 
\cdot \hat{\rho}_{j0j} (t) )  \right.  \nonumber\\
& + & t^{\parallel}_T (k)\ (\hat{\rho}^{\parallel}_j (p) \cdot 
\hat{\rho}^{\parallel}_j (t) )+  t^{\bot}_T (k)(\ (\hat{\rho}^{\bot}_j (p) \cdot
\hat{\rho}^{\bot}_j (t) ) - (\hat{\rho}_{j1j}(p) \cdot \hat{\rho}_{j1j}(t))\ ) 
\nonumber \\
&-& {1\over4} t^T_{LS}(k)\ (\ (\hat{\rho}^{ls}_j (p)\cdot \hat{\rho}_{j0j}(t)) 
+ (\hat{\rho}_{j0j}(p)\cdot \hat{\rho}^{ls}_j (t)) + (\hat{\rho}^{\bot}_j (p)
\cdot \hat{\rho}^l_{jj}(t)) \nonumber \\
&+& (\hat{\rho}^{l}_{jj}(p)\cdot \hat{\rho}^{\bot}_{jj}(t)) + 
(\hat{\rho}^{l\bot}_j (p)\cdot \hat{\rho}_{j1j}(t)) + (\hat{\rho}_{j1j}(p)
\cdot \hat{\rho}^{l\bot}_{j1j}(t))\
) \left. \right\} 
\end{eqnarray}

\noindent
Inserting this decomposition of the $NN$-interaction into the expression 
for the nuclear formfactor
\mbox{$< j_b m_b, J_f M_f \mid \sum_{t} V_{pt} \mid J_i M_i, j_a m_a >$}, we 
obtain \cite{Obsor} formula (\ref{NuclForm}). 

The radial
part of the formfactor can be written in the following form: \\
a) for excitation of \underline{normal} parity states,
\begin{eqnarray}
\lefteqn{F^{lsj}_{j_{a}j_{b}}(\kappa, r_{p},{\partial \over \partial r_{p}} ) = 
\sum_T <T_iM_{T_i}TM_T\vert T_fM_{T_f}><T_bM_{T_b}TM_T\vert T_aM_{T_a}>
{\displaystyle{(-1)^{T_b-T_a}\over \hat{T}_f\hat{T}_a}}\langle{1\over2}
\Vert\tau_p^T\Vert{1\over2}\rangle } \nonumber \\
& & \times {\displaystyle i^j{\hat{s}\over\hat{J}_f}
{2\over\pi}\int_0^{\infty}}dk\ k^2\left\{
\delta_{s0}\delta_{lj}\Biggl[ j_j(kr_p)\ t_{0T}(k)\rho_{j0j,T}(k) 
- {1\over4}t_{LS}^T(k)\rho_{j0j,T}(k)\right. \nonumber \\
& & \times 
\left[(\gamma_a-\gamma_b)j_j(kr_p){1\over \displaystyle k^2r_p}
{\partial \over \partial r_p}+\gamma_a{\displaystyle{dj_j(kr_p)}\over\displaystyle 
dr_p} {1\over \displaystyle k^2r_p}\right.  \\
& &  + {1\over2}\left[j(j+1)-(\gamma_b-\gamma_a)(\gamma_b-\gamma_a+
1)\right]j_j(kr_p){1\over \displaystyle(kr_p)^2}\biggr]\Biggr]  \nonumber \\
& &  - {1\over4}t_{LS}^T(k)\rho_{j1j,T}(k)
\left[\sqrt{j(j+1)}j_j(kr_p){1\over \displaystyle k^2r_p}
{\partial \over \partial r_p}+{\displaystyle{\bigl[(\gamma_b-\gamma_a)
(\gamma_a+\gamma_b+1)-j(j+1)\bigr]}\over\displaystyle 2\sqrt{j(j+1)}}
\right. \nonumber\\
& &  \times \biggl({\displaystyle {j\over 2j+1}}j_{j+1}(kr_p)-
{\displaystyle{j+1\over 2j+1}}j_{j-1}(kr_p)\biggr)\Biggr]
+\delta_{s1}\delta_{lj}j_j(kr_p)\ t_T^{\perp}(k)\rho_{j1j,T}(k) 
\Biggr\} \nonumber
\end{eqnarray}

b) for excitation of \underline{unnatural} parity states \\
\begin{eqnarray}
\lefteqn{F^{lsj}_{j_{a}j_{b}}(\kappa, r_{p},{\partial \over \partial r_{p}} )=
\sum_T <T_iM_{T_i}TM_T\vert T_fM_{T_f}><T_bM_{T_b} TM_T\vert T_aM_{T_a}>
{\displaystyle{(-1)^{T_b-T_a}\over \hat{T}_f\hat{T}_a}}\langle{1\over 2}
\Vert\tau_p^T\Vert{1\over 2}\rangle } \nonumber \\
 & & \times {\displaystyle i^{j+1}{\hat{s}\over\hat{J}_f}
{2\over\pi}\int_0^{\infty}} dk\ k^2\left\{
\delta_{s1}\delta_{l,j-1}j_{j-1}(kr_p)\biggl[-{\displaystyle\sqrt{
{j\over 2j+1}}}t_T^{\Vert}(k)\rho_{j,T}^{\Vert}(k)-{\displaystyle\sqrt{
{j+1\over 2j+1}}}t_T^{\perp}(k)
\rho_{j,T}^{\perp}(k) \right. \nonumber\\
 & & -{1\over2}(\gamma_a+\gamma_b+1-j){\displaystyle
\sqrt{{2j+1\over j+1}}}{1\over4}t_{LS}^T(k)\rho_{j,T}^{\perp}(k)
{1\over \displaystyle kr_p}\biggr] \\
 & & +\delta_{s1}\delta_{l,j+1}j_{j+1}(kr_p)\biggl[
{\displaystyle\sqrt{{j+1\over 2j+1}}}t_T^{\Vert}(k)\rho_{j,T}^{\Vert}(k)-
{\displaystyle\sqrt{{j\over 2j+1}}}t_T^{\perp}(k)\rho_{j,T}^{\perp}(k)
\biggr]\Biggr\} \nonumber
\end{eqnarray}

\noindent
where 
\begin{equation}
\gamma_a =\langle j_{a}m_{a}|\bbox{L}\cdot\bbox{\sigma}|j_{a}m_{a}\rangle =
\left\{ \begin{array}{cc}
    l_a,    & j_a = l_a + \frac{1}{2} \\
-(l_a + 1), & j_a = l_a - \frac{1}{2}
\end{array} \right .
\end{equation}
In the formulas above 
$\rho_{lsj,T}(k) = \langle J_fT_f\Vert \sum_t \hat{\rho}_{lsj} (t)
\tau_t^T\Vert J_iT_i\rangle $ is a complex expression containing
spin-angle reduced matrix elements and one-dimensional integrals over
radial parts of different components of bound and continuum wave functions
and given in Appendix~A. Other densities are the different linear
combinations

\begin{eqnarray}
\rho_{j,T}^{\Vert}(k) &=&{\displaystyle \sqrt{{j\over 2j+1}}}\rho_{j-11j,T}
(k) + {\displaystyle \sqrt{{j+1\over 2j+1}}}\rho_{j+11j,T}(k) \\ 
\rho_{j,T}^{\perp}(k) &=&{\displaystyle \sqrt{{j+1\over 2j+1}}}
\rho_{j-11j,T}(k)
-{\displaystyle \sqrt{{j\over 2j+1}}}\rho_{j+11j,T}(k)
\end{eqnarray}

In radial formfactors we omit the contributions from the current
$\rho_{lj,T}^l(k) = \langle J_fT_f\Vert \sum_t \hat{\rho}^l_{lj} (t)
\tau_t^T\Vert J_iT_i\rangle $ and spin-current 
$\rho_{j,T}^{ls}(k) = \langle J_fT_f\Vert \sum_t \hat{\rho}^{ls}_{j} (t)
\tau_t^T\Vert J_iT_i\rangle $ densities which we did not take into account
in calculations.

The transition densities in coordinate and momentum space are simply connected 
 by
\begin{equation}
\rho_{lsj,T}(r) = {2\over \pi}
\int_0^{\infty} dq\ q^2 j_l(qr) \rho_{lsj,T}(q)
\end{equation}
 

\onecolumn
\small
\newpage
\tableofcontents
\newpage
\listoffigures
\newpage
\listoftables
\normalsize
\newpage
\end{document}